\documentclass[a4paper]{JHEP3}
\usepackage{amsmath}
\usepackage{amsfonts}
\usepackage{bbm}
\usepackage{graphics}
\usepackage{graphicx}
\usepackage{subfigure}
\usepackage{pstricks}

\newcommand{\ben}{\begin{eqnarray}}
\newcommand{\een}{\end{eqnarray}}

\newcommand{\cx}{\mathcal}

\newcommand{\bb}[1][\overline]{1}

\newcommand{\zh}{\hat{z}}

\newcommand{\IP}{{\bf P}}
\newcommand{\IZ}{{\bf Z}}

\newcommand{\ga}{\gamma}

\newcommand{\Si}{\Sigma}

\newcommand{\cxH}{\mathcal{D}}
\def\eqref#1{(\ref{#1})}

\title{Flat Connections in Open String Mirror Symmetry}
\author{Murad Alim$^1$,  Michael Hecht$^2$, Hans Jockers$^3$, Peter Mayr$^2$, Adrian Mertens$^2$ and Masoud Soroush$^3$
\\
\\ $^1$Jefferson Physical Laboratory, Harvard University, Cambridge, MA, 02138, USA
\\ $^2$Arnold Sommerfeld Center for Theoretical Physics, LMU, Theresienstr. 37, D-80333 Munich, Germany
\\ $^3$Bethe Center for Theoretical Physics, Physikalisches Institut
Universit\"{a}t Bonn, Nussallee 12, 53115 Bonn, Germany}

\abstract{We study a flat connection defined on the open-closed deformation space of open string mirror symmetry for type II compactifications on Calabi--Yau threefolds with D-branes. We use flatness and integrability conditions to define distinguished flat coordinates and the superpotential function at an arbitrary point in the open-closed deformation space. Integrability conditions are given for concrete deformation spaces with several closed and open string deformations. We study explicit examples for expansions around different limit points, including orbifold Gromov-Witten invariants, and brane configurations with several brane moduli. In particular, the latter case covers stacks of parallel branes with non-Abelian symmetry.}
\preprint{LMU-ASC 49/11\\BONN-TH-2011-16}

\begin{document}
\section{Introduction}
Mirror symmetry provides the insights and techniques to study the variation of physical structures over their moduli spaces and to perform their exact computation. It  identifies deformation families of the topological string A- and B-models on mirror Calabi--Yau (CY) threefolds $Z$ and $Z^*$. The application of open-string mirror symmetry to $\cx N=1$ supersymmetric compactifications of type II strings on CY threefolds with branes has been pioneered in refs.~\cite{Vafa:1998yp,Kachru:2000ih,Kachru:2000an} and in particular in ref.~\cite{Aganagic:2000gs}, where the authors defined a large class of mirror pairs of brane geometries and obtained the first prediction for Ooguri--Vafa invariants \cite{Ooguri:1999bv} in non-compact geometries from a $B$-model computation. Building on these results  a Hodge theoretic approach to open-string mirror symmetry and the computation of D-brane superpotentials was put forward for branes on local CY in refs.~\cite{Lerche:2002ck,Lerche:2002yw}, for rigid branes on compact CY in refs.~\cite{Walcher:2006rs,Morrison:2007bm} and for branes with open string deformations on compact CY in refs.~\cite{Jockers:2008pe,Alim:2009rf,Alim:2009bx}. This was further studied in refs.~\cite{Jockers:2009mn,Grimm:2009ef,Aganagic:2009jq,Li:2009dz,Alim:2010za}.\footnote{See also refs.~[18--40]
for related work.}

In the Hodge theoretic approach of refs.~\cite{Lerche:2002ck,Lerche:2002yw,Jockers:2008pe,Alim:2009rf,Alim:2009bx}, the open-closed string data are described by the period integrals on a relative cohomology group defined by a divisor $\cxH$ in $Z^*$. The Hodge filtration on the relative cohomology bundle defines a grading which matches the expectations for a $U(1)$ charge in a CFT description, and the Gauss-Manin connection leads to a canonical definition of flat coordinates at grade 1 and potential functions at grade 2. Moreover, the flat coordinates as defined by the Gauss-Manin connection agree with the physical tension of domain walls and an expansion of the appropriate period of grade 2 in terms of these coordinates yields an infinite series, whose coefficient compute -- conjecturally -- the integral Ooguri-Vafa invariants \cite{Ooguri:1999bv} of a D-brane configuration defined by the divisor $\cxH$ and a choice of a D5 charge on $\cxH$, the latter selecting a particular period function. The results obtained from mirror symmetry can be sometimes verified by independent methods, e.g. using localization methods for non-compact CY geometries \cite{Katz:2001vm,Graber:2001dw} or for rigid branes without open string moduli \cite{Walcher:2006rs,pandharipande-2008-21,Krefl:2008sj}. On the other hand there are no independent methods at present to check the predictions of refs.~\cite{Alim:2009rf,Jockers:2009mn,Alim:2009bx,Jockers:2009ti} for Ooguri-Vafa invariants in compact geometries. Moreover  there is as of now no proper understanding of the A model version of the extended Hodge structure and its Gauss-Manin connection, which appear to predict the existence of an extension of the closed string quantum cohomology ring which involves open string states.

In the present work we further elaborate on general aspects of the Gauss-Manin connection on the open-closed deformation space as defined in the B-model, building on the discussion in \cite{Alim:2009bx}. In section 2 we show how the flat structure of the Gauss-Manin connection allows to identify the flat coordinates and the physical couplings everywhere in the deformation space. We apply the technique to investigate the phases of a D-brane on local $\mathbbm{P}^2$ and compute Gromov-Witten invariants at various limit points including the orbifold locus. In section 3 we investigate the integrability conditions on the entries of the connection matrices of the Gauss-Manin connection  for deformation problems with several deformations. These general findings are subsequently studied at the hand of some explicit examples together with their physical interpretation.

\section{Flat structure of $\mathcal{N}=1$ mirror symmetry}

In the setup of refs.~\cite{Lerche:2002ck,Lerche:2002yw,Jockers:2008pe,Alim:2009rf,Alim:2009bx}, the B-model side of open-closed mirror symmetry is described by a relative cohomology group
$H^3(Z^*,\cxH)$ defined on a family of Calabi Yau (CY) threefolds $Z^*$ and a family of holomorphic divisors $\cxH$. The embedding $i:\cxH \hookrightarrow Z^*$ is defined as a complete intersection and depends on complex parameters $\hat{z}_.$. The variation of mixed Hodge structure on the relative cohomology group induced by a variation of the parameters $\hat{z}_.$ is unobstructed. Adding D5-brane charge by turning on flux on a two cycle $C \in H_2(\cxH)$ induces a $\hat{z}$- dependent superpotential if $[C]\in ker(H_2(\cxH) \rightarrow H_2(X))$.

The variation of mixed Hodge structure comes with a flat Gauss-Manin connection which leads to a non-trivial
``$\cx N=1$ special geometry'' of the combined open-closed deformation space of the topological string theory \cite{Mayr:2001xk,Lerche:2002yw,Lerche:2002ck}. In relative cohomology the relative three-form $\underline{\Omega}(z_.,\hat{z}_.)$ acquires a dependence on open string deformations through the presence of the family of divisors and can hence be used to capture the open-closed variational problem.\footnote{For more details on this structure we refer to refs.\cite{Lerche:2002ck,Lerche:2002yw,Jockers:2008pe,Alim:2009rf, Alim:2009bx}.}
To derive a set of differential equations one needs to keep track of exact pieces of the three-forms. Eventually one obtains a Picard-Fuchs like system of differential equations satisfied by the relative period integrals \cite{Lerche:2002ck,Lerche:2002yw,Jockers:2008pe,Alim:2009rf,Alim:2009bx}
\begin{equation}
\label{epf}
\cx L_a \Pi_\Sigma = 0,\qquad \Pi_\Si(z,\zh)=\int_{\ga_\Si} \underline{\Omega}(z_.,\hat{z}_.),\qquad \ga_\Si\in H_3(Z^*,\cxH) \ .
\end{equation}
The system $\{\cx L_a\}$ of linear differential operators which will be central to the following discussion. Its solutions $\Pi_\Si(z,\zh)$ determine the mirror
map and the combined open-closed string superpotential.

In this section we continue the study of the flat structure defined by the Gauss-Manin connection on the relative cohomology bundle, building in particular on ref. \cite{Alim:2009bx}, to which we refer for definitions and examples. We define the flat coordinates and the physical couplings everywhere in the deformation space from the solutions to the differential equations \eqref{epf}. To do this we recall that the variation of mixed Hodge structure describes the variation of the states of the B-model with definite, integral $U(1)$ charge under a deformation. In the CFT, multiplication by a deformation operator of $U(1)$ charge 1 should be described by an upper triangular matrix acting on the state vector, reflecting multiplication in the chiral ring of the CFT. We refer to \cite{Lerche:1991wm,Hosono:1996jv,Hosono:1995bm,Bershadsky:1993cx} for a detailed discussion of this issues in the closed string theory and concentrate on the new aspects that arise from the inclusion of open string deformations. Geometrically the insertion of deformation operators corresponds to taking derivatives with respect to the deformation parameters.

\subsection{Deformation families with one closed and one open modulus}
For simplicity, we start with a qualitative discussion for the case of one closed and one open modulus, which describes e.g. the brane geometry on the quintic considered in \cite{Alim:2009bx}. Let $z,\hat{z}$ denote arbitrary local algebraic complex structure coordinates parameterizing closed and open string deformations, respectively. Starting from the relative holomorphic $(3,0)$ form $\underline{\Omega}(z_{.})$ we consider a basis for the Hodge filtration
\begin{equation}
\begin{array}{ccccccccc}
\vec{\underline{\Omega}}(z_.) &=&
(\underline{\Omega}(z_.)&
 \partial_{z} \underline{\Omega}(z_.) &\partial_{\hat{z}} \underline{\Omega}(z_.)&
 \partial_{z}^2 \underline{\Omega}(z_.)&\partial_{z} \partial_{\hat{z}} \underline{\Omega}(z_.)&
\partial_{z}^3\underline{\Omega}(z_.) &\partial_{z}^2 \partial_{\hat{z}} \underline{\Omega}(z_.))^T
\end{array}
\end{equation}
We assume that all other multi-derivatives up to degree 3 can be expressed as linear combinations of these using  the Picard-Fuchs (PF) operators. These then allow us to express the derivatives of $\vec{\Omega}(z_.)$ in this basis. In particular we obtain
\begin{equation}
(\partial_z - M_z) \vec{\underline{\Omega}}(z_.)=0 \, ,\quad (\partial_{\hat{z}} - M_{\hat{z}}) \vec{\underline{\Omega}}(z_.)=0 \, ,
\end{equation}
with connection matrices $M_z,M_{\hat{z}}$. These will not have the upper triangular form expected from a CFT description, however. To find the coordinates in terms of which the connection matrices have upper triangular form, we make the following ansatz:
\begin{equation}
\vec{\underline{\Omega}}'(t_.)=
\begin{array}{cccc}
(\underline{\Omega'}(t) &
\partial_{t_.} \underline{\Omega'}(t) &
C_t^{(1)}(t_.)^{-1}\partial_{t_.}^2 \underline{\Omega'}(t_.)&
C_t^{(2)}(t_.)^{-1}\partial_{t} (C_t^{1}(t_.)^{-1}\partial_{t_.}^2 \underline{\Omega'}(t_.) )^T \,
\end{array}
\end{equation}
where
\begin{equation} \label{omegat}
 \underline{\Omega}'(t_.) =\frac{\underline{\Omega}(z_.(t_.))}{\omega_0(t_.)}\, , \quad C_t^{(1)}=\left( \begin{array}{cc} C &W_{tt} \\0& W_{t\hat{t}} \end{array}\right)
\, , \quad C_t^{(2)}=\left( \begin{array}{cc} -1 &\mu_t  \\0& \rho_{t} \end{array}\right) \, ,
\end{equation}
and
\begin{equation}
\partial_{t_.} \underline{\Omega'}(t_.) = \left( \begin{array}{c} \partial_{t} \underline{\Omega'}(t) \\ \partial_{\hat{t}} \underline{\Omega'}(t_.) \end{array}\right) \, ,\quad
\partial_{t_.}^2 \underline{\Omega'}(t_.) = \left( \begin{array}{c} \partial_{t}^2 \underline{\Omega'}(t) \\ \partial_t \partial_{\hat{t}} \underline{\Omega'}(t_.) \end{array}\right) \, .
\end{equation}
Here $t_.$ stands collectively for the searched for flat coordinates $t$ and $\hat{t}$ and the ansatz for the $t_.$-dependent functions is motivated by the discussion in \cite{Alim:2009bx}. These functions can be computed from the complex structure variation encoded in the Picard-Fuchs equations as follows. We first determine the matrix $A(t_.)$ relating the two bases
$$ \vec{\underline{\Omega}}'(t_.)=A(t_.) \vec{\underline{\Omega}}(z_.) \, .$$
The new connection matrices $M_{t_.}$ can be computed out of $M_{z_.}$ and $A$:
\begin{equation}
M_{t_.} = \partial_{t_.} A \cdot A^{-1} + \sum_{z_.} \frac{\partial z_.}{\partial t_.} A \cdot M_{z_.} A^{-1}\,.
\end{equation}
The entries of the matrices $M_{t_.}$ which are not upper triangular are differential equations for the unknown functions. Setting these entries to zero, reformulates the Picard-Fuchs equations into differential equations for the desired quantities. E.g., the normalization factor $\omega_0(t)$ is determined to be proportional to a particular solution of the Picard-Fuchs equations. In addition one obtains differential equations for the mirror maps $t_.(z_.)$ and the functions $C^{(q)}(t_.)$. Solving these equations, the variation of mixed Hodge structure becomes, in the new coordinates
\begin{equation}
(\partial_{t_.} - M_{t_.}) \vec{\underline{\Omega}}(t_.)=0\,,
\end{equation}
with matrices of the general form \cite{Alim:2009bx}:
\begin{equation} \label{Mt1}
M_t= \left(\begin{array}{ccccccc}  0 &1&0&0&0&0&0  \\
0&0&0&C&W_{tt}&0&0 \\
0&0&0&0&W_{t\hat{t}}&0&0 \\
0&0&0&0&0&-1&\mu_t \\
0&0&0&0&0&0&\rho_t \\
0&0&0&0&0&0&0\\
0&0&0&0&0&0&0
\end{array} \right) \, ,\quad
M_{\hat{t}}= \left(\begin{array}{ccccccc}  0 &0&1&0&0&0&0  \\
0&0&0&0&W_{t\hat{t}}&0&0 \\
0&0&0&0&W_{\hat{t}\hat{t}}&0&0 \\
0&0&0&0&0&0&\mu_{\hat{t}} \\
0&0&0&0&0&0&\rho_{\hat{t}} \\
0&0&0&0&0&0&0\\
0&0&0&0&0&0&0
\end{array} \right) \, .
\end{equation}

\subsection{A subsystem capturing open string deformations}
In flat coordinates the Picard-Fuchs equations become very simple
\begin{equation}\label{pfflat}
\partial_{t} (C_t^{(2)})^{-1\phantom{t^c} t^b}_{\phantom{-1} t^c} \partial_{t} (C_t^{(1)})^{-1 \phantom{t^b} t^a}_{\phantom{-1} t^b} \partial_{t^a} \partial_{t} \underline{\Omega}'(t_.) =0\, ,
\end{equation}
where $t^a,t^b,t^c$ run over $t$ and $\hat{t}$, $a$ and $b$ are summed over and c is a free index giving two equations. For $t^c=t$ we obtain an equation
\begin{equation}
\partial_t^2 C^{-1} \partial_t^2 \underline{\Omega}'(t_.) + \mathcal{L}_{bdry} \partial_{\hat{t}} \underline{\Omega}'(t_.) =0 \, ,
\end{equation}
where the first part is the Picard-Fuchs equation for the closed string geometry without open string deformations. The equation for $t^c=\hat{t}$ reads
\begin{equation} \label{pfflathat}
\partial_t\, \rho_t^{-1} \partial_t \, W_{t\hat{t}}^{-1} \partial_t  \, \partial_{\hat{t}} \underline{\Omega}'(t_.)=0\, .
\end{equation}
Now $\partial_{\hat{t}} \underline{\Omega}'(t_.)$ corresponds in relative cohomology to the two form $\omega'(t_.)\in H^{2,0}(\cxH)$, capturing the variation of the Hodge structure on the divisor $\cxH$ \cite{Alim:2009bx}. A second order operator, expressing the linear dependence of the elements of charge 2, can be read off the connection matrices (\ref{Mt1}):
\begin{equation}
\left( W_{t\hat{t}} \partial_{\hat{t}}-W_{\hat{t}\hat{t}} \partial_{t}\right) \, \omega'(t_.) =0 \, .
\end{equation}
This equation is solved by $\omega'=\omega'(W_{\hat{t}})$.  Rewriting the operator (\ref{pfflathat}) in terms of $\omega'(W_{\hat{t}})$ finally gives
\begin{equation} \label{pfflatK3}
\partial_{W_{\hat{t}}}^3 \,  \omega'(W_{\hat{t}}) =0 \, .
\end{equation}
Here we have inserted $\rho= a W_{\hat{t}}+b$, with $a,b$ some constants, which follows from the integrability conditions, as shown in ref.~\cite{Alim:2009bx}. We recognize equation (\ref{pfflatK3}) as describing the Picard-Fuchs equation for a $K3$ manifold, expressed in terms of the flat K3 coordinate $t_{K3}=W_{\hat{t}}$. In the full deformation problem, this parameter captures the variation of the relative periods under a deformation in the open string sector.

\subsection{Connection matrices in arbitrary coordinates}
The special form of the variation of mixed Hodge structure above is due to the existence of the flat Gauss-Manin connection on the relative cohomology group defined in ref.~\cite{Alim:2009bx}. The flat coordinates, which are the ones needed for mirror symmetry, are merely a special choice of coordinates to describe the problem. The upper triangular structure of the connection matrices can be traced back to Griffiths transversality of the Gauss-Manin connection on the filtration spaces $\nabla F^q \subset F^{q-1}$. In the following we will outline how the structure can be seen in arbitrary coordinates. We will do this by giving the expressions for the covariant derivatives $\nabla_{z_.}$ in arbitrary coordinates.

\begin{itemize}
\item {\it Connection for the line bundle} \\
Going from algebraic coordinates to flat coordinates we chose a specific normalization of the $(3,0)$ form by normalizing it with $\omega_0(t)$. This reflects the fact that the holomorphic $(3,0)$ form is a section of a line bundle $\mathcal{L}\rightarrow \mathcal{M}$ over the deformation space. The derivative in coordinates where we haven't chosen a specific normalization should hence contain a connection factor correcting this

$$ \partial_t \Omega'(t_.) \leftrightarrow  \left( \partial_{z_.} - \partial_{z_.} \log{\omega_0(t_.(z_.))} \right) \Omega(z_.)$$

\item{\it Connection on the cotangent space}\\
The reason the mirror maps are called flat coordinates is because the connection  $\Gamma_{t_.}$ of the cotangent bundle over the deformation space $\mathcal{T}^* \mathcal{M}$ vanishes in these coordinates (see ref.~\cite{Bershadsky:1993cx}). It is now possible to compute what the connection in arbitrary coordinates should be without referring to the metric just by transforming the vanishing connections $\Gamma_{t_.}$ to other coordinates. This gives
\begin{equation}
\Gamma_{ij}^k= \frac{\partial z^k}{\partial t^a} \frac{\partial^2 t^a}{\partial z^i \partial z^j}
\end{equation}
\end{itemize}
We hence have found the explicit form of $\nabla_{i}$ in arbitrary coordinates. It contains the connections of the line bundle and the cotangent bundle
\begin{equation}
\nabla_i = \partial_{i} -\Gamma_i -\partial_i \log{\omega_0} \, , \quad \partial_i=\frac{\partial}{\partial z^i}\, ,\quad i=1,\dots,h^{2,1}(Z^*)+ h^{2,0}(\cxH)\, .
 \end{equation}
Flat coordinates for the open-closed deformation space are defined to be the ones in which both connection factors vanish. With this definition, the flat coordinates are the open-closed string generalization of the closed string canonical coordinates defined in ref.~\cite{Bershadsky:1993cx}, after taking the holomorphic limit.

\subsection{Example: local $\mathbbm{P}^2$ with one open modulus}

As an illustration of the techniques discussed so far we study an example of  the non-compact geometry of local $\mathbbm{P}^2$ with one closed and one open string modulus. At the combined large volume open-closed expansion locus one can extract the Ooguri-Vafa invariants \cite{Ooguri:1999bv}, first computed in ref.\cite{Aganagic:2001nx}. This was also the main example of the Hodge-theoretic approach of  \cite{Lerche:2001cw,Lerche:2002ck,Lerche:2002yw}.\footnote{See \cite{Lerche:2003hs} for an excellent review and a clear exposition of the use of the Gauss-Manin system for finding flat coordinates in the large volume phases.}

With the uprise of orbifold Gromov-Witten invariants\footnote{See \cite{Bouchard:2007nr,Brini:2010sw} and references therein for the mathematical definitions and refs.\cite{Aganagic:2006wq,Bouchard:2007ys,Bouchard:2008gu,Brini:2008rh,Alim:2008kp,Brini:2011ij} for a study of this expansion locus in the topological string.}, limit points different from the large complex structure point have gained considerable interest, as one may now check the predictions of mirror symmetry explicitely at a point in the deformation space different from the large volume limit of the A-model. In the following we use the variation of mixed Hodge structures and the flatness of the Gauss-Manin connection to study the flat coordinates of $\mathcal{N}=1$ mirror symmetry and the three-point functions globally in the deformation space.

The open-closed Picard-Fuchs system is read off from the charge vectors:

\begin{equation}
\begin{array}{c|cccccc}
&a_0&a_1&a_2&a_3&a_4&a_5\\
\hline
l^1&-3&1&1&1&0&0\\
l^2&1&-1&0&0&1&-1
\end{array}
\end{equation}

\noindent
The algebraic coordinates

\begin{equation}
z= - \frac{a_1 a_2 a_3}{a_0^3}\, ,\quad \hat{z}=-\frac{a_0 a_4}{a_1 a_5} \, ,
\end{equation}
are centered around the large complex structure point describing the outer phase of the geometry in refs.\cite{Aganagic:2001nx,Lerche:2001cw,Lerche:2002ck,Lerche:2002yw}. The Picard-Fuchs operators correspond to the GKZ operators which can be read off from the charge vectors $l_1,l_2$ and $l_3=l_1-l_2$, these are:
\begin{eqnarray} \label{pflocalp2}
\mathcal{L}_1 &=& (\theta-\hat{\theta}) \theta^2 - z \prod_{i=0}^{2} (3\theta-\hat{\theta}+i)=\mathcal{L}_{bulk} + \mathcal{L}_{bdry}\, \hat{\theta} \, ,\nonumber\\
\mathcal{L}_2 &=& \left( (3 \theta - \hat{\theta}) + \hat{z} (\theta- \hat{\theta})  \right)  \hat{\theta}=\mathcal{L}'_2 \,\hat{\theta} \, ,\nonumber\\
\mathcal{L}_3 &=& \left( \theta^2 + z \hat{z}  \prod_{i=0}^{1} (3\theta-\hat{\theta}+i) \right) \hat{\theta}=\mathcal{L}'_3 \,\hat{\theta}\, ,\nonumber\\
\mathcal{L}_{bulk} &=& \theta^3 - z \prod_{i=0}^{2} (3\theta+i) \, ,\nonumber\\
\mathcal{L}_{bdry} &=& \theta^2- z \left( (3 \theta -\hat{\theta}+1)(3 \theta -\hat{\theta}+2)+3 \theta(3 \theta -\hat{\theta}+2)+3 \theta(3 \theta +1)\right)\, .
\end{eqnarray}
The discriminants are
\begin{equation}
\Delta_1 = 1-27 z \, ,\quad \Delta_2= (1+\hat{z})^2+ 4 z \hat{z}^3\, ,  \quad \Delta_3= 1+\hat{z} \, .
\end{equation}
The solution space of these operators is five dimensional, as expected, with a basis given by the relative periods $\Pi_i(z,\hat{z}), i=0,\dots,4$ (see \eqref{epf}).  Three of the solutions $\Pi_{i}(z)\,, i=0,1,2$ are $\hat{z}$ independent and correspond to solutions of $\mathcal{L}_{bulk}$, which have been studied in detail in ref. \cite{Diaconescu:1999dt}.

\subsubsection{Subsystem}
It is useful to connect the other two solutions to the periods of the subsystem of periods defined on the divisor $\cxH$ discussed above and in refs.\cite{Jockers:2008pe,Alim:2009rf, Alim:2009bx}. In particular we can use the understanding of the phases of the deformation space of this subsystem to discuss the combined phases of open-closed mirror symmetry.

To discuss the structure of the open-string sector we write the corresponding solutions of the PF system in terms of $ \pi_{a}(z,\hat{z})= \hat{\theta} \,\Pi_{a+3} (z,\hat{z}),\, a=0,1$, which satisfy the equations:
$$ \mathcal{L}'_i \, \pi_{a} (z,\hat{z}) =0\,, \quad  i=2,3\, , \quad a=0,1. $$
We note that the kernel of the first order operator $\mathcal{L}'_2$ allows us to redefine the coordinates. In terms of the new coordinate
$$ z_{\cxH} =-\frac{z \hat{z}^3}{(1+\hat{z})^2} \ , $$
the operator $\mathcal{L}'_3$ becomes
\begin{equation}
\mathcal{L}'_3=\theta^2-2 z_{\cxH} \theta (2 \theta+1) \, ,\quad \theta=z_{\cxH} \frac{d}{d z_{\cxH}} \,,
\end{equation}
which is the Picard-Fuchs operator of the variation of Hodge structure associated to the cotangent bundle $\mathcal{O}(-2) \rightarrow \mathbbm{P}^1$, which is described by the charge vector
\begin{equation}
\begin{array}{c|ccc}
&a_0&a_1&a_2\\
l_{\cxH}&-2&1&1\,\\
\end{array}
\end{equation}
giving the algebraic coordinate
 $$ \quad z_{\cxH}=\frac{a_1 a_2}{a_0^2}\,.$$
The discriminant is $\Delta=1-4 z_{\cxH}$. The solutions of this equation are the constant $\pi_0(z_{\cxH})=1$ and the mirror map which is given by
\begin{equation} \label{O2map}
\pi_1(z_{\cxH})= t^{lcs}_{\cxH}(z_{\cxH})= \log \frac{1-\sqrt{1-4 z_{\cxH}}}{1+\sqrt{1-4 z_{\cxH}}} \, .
\end{equation}
Different phases of the open string sector correspond to different phases of this subsystem which correspond to expansion loci around singular points in complex structure space. We identify a large complex structure expansion point, an orbifold point and the discriminant locus.
\begin{itemize}
\item The large complex structure point is described by the coordinate $z_{\cxH}$ giving the expansion of (\ref{O2map})
$$ t^{lcs}_{\cxH} (z_{\cxH})= \log z_{\cxH} + 2z_{\cxH} + 3 z_{\cxH}^2+\dots$$
\item In the vicinity of small $x_{\cxH}=z_{\cxH}^{-1/2}$ the deformation space describes (the mirror of) an $\mathbbm{Z}_2$ orbifold singularity,  with the mirror map
$$ t^{orb}_{\cxH}(x_{\cxH})= \pi - i t^{lcs}_{\cxH}= x_{\cxH}+\frac{1}{24} x_{\cxH}^3+\dots$$
\item The vicinity of small $y=\sqrt{\Delta}$  describes a conifold singularity, the mirror map is given by
$$ t^{con}_{\cxH}(y_{\cxH}) = -\frac{1}{2} t^{lcs}_{\cxH} = \sum_{i=0}^{\infty} \frac{y_{\cxH}^{2i+1}}{2i+1}$$
\end{itemize}


\subsubsection{Open-closed phases}
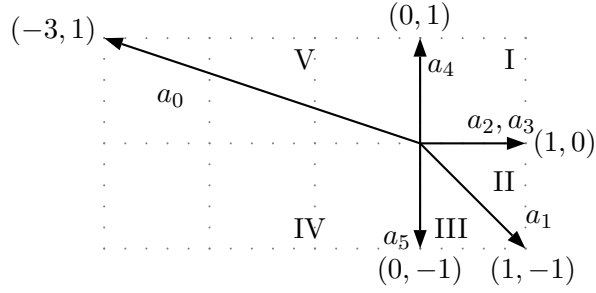
\begin{figure}[t!]
\centering
\begin{minipage}{0.7\textwidth}
\centering

\psset{unit=1.4cm}
\psset{griddots=0,gridlabels=0,subgriddiv=5}
\begin{pspicture}(-3,-1.5)(1,1)
\psgrid[griddots=5,subgriddiv=0,gridcolor=gray]
\psline[arrows=->,arrowsize=5pt,arrowinset=0](0,0)(1,0)
\psline[arrows=->,arrowsize=5pt,arrowinset=0](0,0)(0,1)
\psline[arrows=->,arrowsize=5pt,arrowinset=0](0,0)(1,-1)
\psline[arrows=->,arrowsize=5pt,arrowinset=0](0,0)(0,-1)
\psline[arrows=->,arrowsize=5pt,arrowinset=0](0,0)(-3,1)

\rput[l](1.07,0){$(1,0)$}
\rput[b](0,1.07){$(0,1)$}
\rput[r](-3.07,1.07){$(-3,1)$}
\rput[t](1.07,-1.07){$(1,-1)$}
\rput[t](0,-1.07){$(0,-1)$}
\rput[bl](-2.5,0.35){$a_0$}
\rput[bl](1,-0.8){$a_1$}
\rput[br](1.1,0.09){$a_2,a_3$}
\rput[tl](0.07,0.8){$a_4$}
\rput[bl](-0.35,-1){$a_5$}
\rput[tr](0.9,0.9){I}
\rput[br](0.9,-0.45){II}
\rput[br](0.45,-0.9){III}
\rput[bl](-1.2,-0.9){IV}
\rput[tl](-1.2,0.9){V}
\end{pspicture}
\caption{Secondary fan for open-closed moduli space of local $\mathbbm{P}^2$.}
\label{fig_p2}
\end{minipage}
\end{figure}

Having discussed the phases of the subsystem which captures open string data we now turn to the analysis of the full open-closed deformation space for this problem. After relating the previous discussion to ref.~\cite{Lerche:2001cw} where the subsystem was observed, we will use the techniques to describe other regions in moduli space. The different phases of the open-closed deformation space are encoded in terms of the homogeneous coordinates in the secondary fan, which is depicted in fig.~\ref{fig_p2}.

\subsubsection*{\it Large Volume}
This problem allows two large volume descriptions related by a flop transition leading to the outer and the inner phase \cite{Aganagic:2001nx,Lerche:2001cw}. In the following we will only briefly recall the flat coordinates for these phases and present the Gauss-Manin matrices which already appeared in ref.~\cite{Lerche:2003hs}.

\begin{itemize}
\item Outer phase (\emph{Region I}): \\
Good coordinates in this patch are given by $z,\hat{z}$. Bringing the Gauss-Manin matrices $M_z$ and $M_{\hat{z}}$ to upper triangular form leads to the mirror maps \cite{Lerche:2001cw}
\begin{equation}
t(z)= \log z + 3 A(z) \, ,\quad \hat{t}(z,\hat{z})= \log \hat{z}-A(z)\, ,\quad A(z)=\sum_{n=0}^{\infty}\frac{(3n-1)!}{(n!)^3} z^n \, .
\end{equation}
The solution corresponding to the superpotential satisfies
\begin{equation}
\hat{\theta} \mathcal{W}(z,\hat{z}) = t^{lcs}_{\cxH} \left( -\frac{z \hat{z}^3}{(1+\hat{z})^2}\right) \,.
\end{equation}
\item Inner Phase (\emph{Region II}): \\
The coordinates parameterizing this flopped region of moduli space are given by $z_1= z \hat{z}$ and $z_2=1/\hat{z}$ with the mirror maps
\begin{equation}
t_1(z_1,z_2)= \log z_1 + 2 A(z_1 z_2) \, ,\quad t_2(z_1,z_2)= \log z_2+A(z_1 z_2)\,  .
\end{equation}
And similarly the superpotential satisfies
\begin{equation}
(\theta_1-\theta_2) \mathcal{W}(z_1,z_2) = t^{lcs}_{\cxH} \left( -\frac{z_1}{(1+z_2)^2}\right) \,.
\end{equation}
\end{itemize}
We refer to ref.\cite{Lerche:2001cw} for the superpotential expansion in algebraic and in flat coordinates.

To obtain the Gauss-Manin  connection matrices we take the vector spanning the Hodge filtration to be:
$$ \vec{\Omega} = (\Omega,\partial_z \Omega,\partial_{\hat{z}} \Omega,\partial^2_z \Omega,\partial_z \partial_{\hat{z}} \Omega) \,.$$
We then translate the Picard-Fuchs equations into the equations
\begin{equation}
(\partial_z -M_z) \vec{\Omega} = 0 \, ,\quad (\partial_{\hat{z}} -M_{\hat{z}}) \vec{\Omega}=0  \, ,
\end{equation}
where the connection matrices are given by

\begin{equation}
M_{z}=\left(
\begin{array}{ccccc}
 0 & 1 & 0 & 0 & 0 \\
 0 & 0 & 0 & 1 & 0 \\
 0 & 0 & 0 & 0 & 1 \\
 0 & \frac{-1+60 z}{z^2 \Delta_1} & -\frac{2 \hat{z}}{z^2 \Delta_1} & -\frac{3-108 z}{z\Delta_1}
   & \frac{-\hat{z} \left((22 z+2) \hat{z}^4+(18 z+17) \hat{z}^3+37 \hat{z}^2+31 \hat{z}+9\right)}{z \Delta_1 \Delta_2
   \Delta_3} \\
 0 & 0 & 0 & 0 & -\frac{6 z \hat{z}^3+\hat{z}^2+2 \hat{z}+1}{z \Delta_2}
\end{array}
\right)
\end{equation}

\begin{equation}
M_{\hat{z}}=\left(
\begin{array}{ccccc}
 0 & 0 & 1 & 0 & 0 \\
 0 & 0 & 0 & 0 & 1 \\
 0 & 0 & -\frac{1}{\hat{z}} & 0 & \frac{z (\hat{z}+3)}{\hat{z} \Delta_3} \\
 0 & 0 & 0 & 0 & -\frac{6 z \hat{z}^3+\hat{z}^2+2 \hat{z}+1}{z \Delta_2} \\
 0 & 0 & 0 & 0 & -\frac{6 z \hat{z}^4+10 z \hat{z}^3+\hat{z}^3+3 \hat{z}^2+3 \hat{z}+1}{\hat{z} \Delta_3 \Delta_2}
\end{array}
\right)
\end{equation}
with
\begin{equation}
\Delta_1 = 1-27 z \, ,\quad \Delta_2= (1+\hat{z})^2+ 4 z \hat{z}^3\, ,  \quad \Delta_3= 1+\hat{z} \, .
\end{equation}

\subsubsection*{\it Mixed phase (Region V)}
We consider the region in moduli space in the vicinity of the algebraic coordinates
\begin{equation}
x= -\frac{a_0}{(a_1 a_2 a_3)^{1/3}} = z^{-1/3} \, ,\quad \tilde{z}= \frac{(a_2 a_3)^{1/3}}{a_1^{2/3}} \left(-\frac{a_4}{a_5} \right) = -z^{1/3} \hat{z}\, ,
\end{equation}
where the first coordinate is just the algebraic coordinate in the vicinity of the orbifold singularity in the closed string moduli space \cite{Diaconescu:1999dt,Aganagic:2006wq}.
The coordinate of the subsystem is found to be $z_{\cxH} = \tilde{z}^3(1-x \tilde{z})^{-2}$ , which shows that the subsystem is still at the large volume expansion locus.  We find the following flat coordinates:
\begin{eqnarray}
t(x)&=&x+\frac{1}{648}  x^4+\frac{4  x^7}{229635}+\frac{49
   x^{10}}{159432300}+ \dots \nonumber\\
\tilde{t}(\tilde{z})&=& \log{\tilde{z}} \, , \quad  \tilde{q}=e^{\tilde{t}} \, ,
\end{eqnarray}
The Superpotential in this region corresponds to the solution of the Picard-Fuchs system satisfying
\begin{equation}
\tilde{\theta}_{\tilde{z}} \mathcal{W} = t^{lcs}_{\cxH} \left(\tilde{z}^3(1-x \tilde{z})^{-2} \right)\, .
\end{equation}
The flat coordinate expansion of the superpotential solution reads
\begin{eqnarray}
\mathcal{W}(t,\tilde{q})&=&\frac{3}{2} \tilde{t}^2 + 2 \tilde{q} t+\frac{2 \tilde{q}^3}{3}+\frac{1}{2}
   \tilde{q}^2 t^2+\left(\tilde{q}^4 t-\frac{\tilde{q} t^4}{324}\right)
+\left(\frac{\tilde{q}^6}{2}+\frac{2 \tilde{q}^3 t^3}{9}\right)+   \left(\frac{6 \tilde{q}^5 t^2}{5}-\frac{\tilde{q}^2 t^5}{648}\right) \nonumber\\&+&\frac{\left(3149280 \tilde{q}^7 t+226800 \tilde{q}^4 t^4-29 \tilde{q}
   t^7\right)}{1837080}+\frac{1}{972}  \left(720 \tilde{q}^9+1296 \tilde{q}^6 t^3-\tilde{q}^3 t^6\right)+\dots
\end{eqnarray}
This is the region in moduli space corresponding to the analysis in ref.~\cite{Bouchard:2008gu}. It was suggested in that paper to extract orbifold disk invariants from the flat coordinate expansion of the superpotential  using the prescription:
\begin{equation}
\mathcal{W}(t,\tilde{q})= \sum_{i,j} \, \frac{1}{j!}\,  N_{i,j} \tilde{q}^i \,  t^j \,,
\end{equation}
where the classical term $\frac{3}{2} \tilde{t}^2$ is not taken into account. Using this prescription we extract the following $N_{i,j}$ from $\frac{1}{2} \mathcal{W}(t,\tilde{q})$:
\begin{equation}
\begin{array}{|c|ccccccc|}
\hline
&&&&&&&\\
j \setminus i&1&2&3&4&5&6&7\\
\hline
&&&&&&&\\
0 & 0 & 0 & \frac{1}{3} & 0 & 0 & \frac{1}{4} & 0 \\
 1 & 1 & 0 & 0 & \frac{1}{2} & 0 & 0 & \frac{6}{7} \\
 2 & 0 & \frac{1}{2} & 0 & 0 & \frac{6}{5} & 0 & 0 \\
 3 & 0 & 0 & \frac{2}{3} & 0 & 0 & 4 & 0 \\
 4 & -\frac{1}{27} & 0 & 0 & \frac{40}{27} & 0 & 0 & \frac{154}{9} \\
 5 & 0 & -\frac{5}{54} & 0 & 0 & \frac{206}{45} & 0 & 0 \\
 6 & 0 & 0 & -\frac{10}{27} & 0 & 0 & \frac{160}{9} & 0 \\
 7 & -\frac{29}{729} & 0 & 0 & -\frac{1432}{729} & 0 & 0 & \frac{19586}{243}\\
&&&&&&&\\
\hline
\end{array}
\end{equation}
which agrees with the invariants given in ref.~\cite{Bouchard:2008gu} up to minus signs which are due to the inclusion of a minus sign in our definition of the algebraic orbifold coordinate.

\subsubsection*{\it Orbifold (Region IV)}
We consider the region in moduli space in the vicinity of the algebraic coordinates
\begin{equation}
x= -\frac{a_0}{(a_1 a_2 a_3)^{1/3}} = z^{-1/3} \, ,\quad \hat{x}= \frac{a_1^{1/3}}{(a_2 a_3)^{1/6}} \left(-\frac{a_5}{a_4} \right)^{1/2} = (-z \hat{z}^3)^{-1/6}\, ,
\end{equation}
where the first coordinate is again the algebraic coordinate in the vicinity of the closed string orbifold locus.
The coordinate of the subsystem is found to be $x_{\cxH} = \hat{x}(-x+\hat{x}^2)$ and hence the proposed coordinates are good coordinates in the vicinity of the orbifold expansion locus of the subsystem.  We consider the combination of the two orbifold expansion loci to be the ``orbifold" locus of open-closed mirror symmetry and find the coordinate transformations that flatten the Gauss-Manin matrices $M_x$ and $M_{\hat{x}}$. The latter are obtained by considering
the vector spanning the filtration to be
$$ \vec{\Omega} = (\Omega,\partial_x \Omega,\partial_{\hat{x}} \Omega,\partial^2_x \Omega,\partial_x \partial_{\hat{x}} \Omega) \,.$$
We then translate the Picard-Fuchs equations into the equations
\begin{equation}
(\partial_x -M_x) \vec{\Omega} = 0 \, ,\quad (\partial_{\hat{x}} -M_{\hat{x}}) \vec{\Omega}=0  \, ,
\end{equation}
where the connection matrices are given by
\begin{equation}
M_x=
\left(
\begin{array}{ccccc}
 0 & 1 & 0 & 0 & 0 \\
 0 & 0 & 0 & 1 & 0 \\
 0 & 0 & 0 & 0 & 1 \\
 0 & -\frac{x}{\Delta_1} & 0 & -\frac{3 x^2}{\Delta_1} & \frac{\hat{x} \left(x^2 \hat{x}^4+18 \hat{x}^2-x
   \left(\hat{x}^6+8\right)\right)}{2 \left(\Delta_1\right) \left(\Delta_2\right)} \\
 0 & 0 & 0 & 0 & \frac{\hat{x}^2 \left(\hat{x}^2-x\right)}{\Delta_2}
\end{array}
\right) \, ,
\end{equation}

\begin{equation}
M_{\hat{x}}=
\left(
\begin{array}{ccccc}
 0 & 0 & 1 & 0 & 0 \\
 0 & 0 & 0 & 0 & 1 \\
 0 & 0 & -\frac{1}{\hat{x}} & 0 & \frac{x}{\hat{x}}-3 \hat{x} \\
 0 & 0 & 0 & 0 & \frac{\hat{x}^2 \left(\hat{x}^2-x\right)}{\Delta_2} \\
 0 & 0 & 0 & 0 & -\frac{\hat{x} \left(3 \hat{x}^4-4 x \hat{x}^2+x^2\right)}{\Delta_2}
\end{array}
\right) \, ,
\end{equation}
with
\begin{equation}
\Delta_1= x^3 -27 \, ,\quad \Delta_2= \hat{x}^6-2 x \hat{x}^4+x^2 \hat{x}^2-4 \, .
\end{equation}

The superpotential in this case is a solution of the Picard-Fuchs equations (\ref{pflocalp2}) translated to the coordinates $x$ and $\hat{x}$ and furthermore satisfies
$$ \hat{\theta}_{\hat{x}} \mathcal{W}(x,\hat{x}) = t^{orb}_{\cxH}\left(  \hat{x}(-x+\hat{x}^2) \right) $$
We find the following flat coordinates:
\begin{eqnarray}
t(x)&=&x+\frac{1}{648}  x^4+\frac{4  x^7}{229635}+\frac{49
   x^{10}}{159432300}+ \dots \nonumber\\
\hat{t}(\hat{x})&=& \log{\hat{x}} \, , \quad  \hat{q}=e^{\hat{t}} \, ,
\end{eqnarray}
and the superpotential in these coordinates is up to normalization
\begin{eqnarray}
\mathcal{W}(t,\hat{q})=&&  \hat{q} t-\frac{\hat{q}^3}{3}-\frac{1}{648} \left(\hat{q} t^4\right)+\frac{1}{72}  \hat{q}^3 t^3-\frac{1}{40}  \left(\hat{q}^5
   t^2\right)+ \left(\frac{\hat{q}^7 t}{56}-\frac{29 \hat{q} t^7}{3674160}\right)+\left(-\frac{\hat{q}^9}{216}-\frac{\hat{q}^3
   t^6}{15552}\right) \nonumber\\
&&+\frac{263  \hat{q}^5 t^5}{259200}+\left(-\frac{\hat{q}^7 t^4}{36288}-\frac{6607 \hat{q}
   t^{10}}{71425670400}\right)+\dots
\end{eqnarray}
Following the closed string multi-moduli orbifold case\footnote{See \cite{Brini:2008rh,Alim:2008kp} and references therein.}, we extract the orbifold invariants from the expression
\begin{equation}
\mathcal{W}(t,\hat{q})= \sum_{i,j} \, \frac{1}{i!\,j!}\,  N_{i,j} \hat{q}^i \,  t^j \,.
\end{equation}

\begin{equation} \label{orbinv}
\begin{array}{|c|ccccccccccc|}
\hline
&&&&&&&&&&&\\
j \setminus i&0&1&2&3&4&5&6&7&8&9&10\\
\hline
&&&&&&&&&&&\\
 0&0 & 0 & 0 & -2 & 0 & 0 & 0 & 0 & 0 & -1680 & 0 \\
 1&0 & 1 & 0 & 0 & 0 & 0 & 0 & 90 & 0 & 0 & 0 \\
 2&0 & 0 & 0 & 0 & 0 & -6 & 0 & 0 & 0 & 0 & 0 \\
 3&0 & 0 & 0 & \frac{1}{2} & 0 & 0 & 0 & 0 & 0 & 11340 & 0 \\
 4&0 & -\frac{1}{27} & 0 & 0 & 0 & 0 & 0 & -\frac{1225}{3} & 0 & 0 & 0 \\
 5&0 & 0 & 0 & 0 & 0 & \frac{263}{18} & 0 & 0 & 0 & 0 & 0 \\
 6&0 & 0 & 0 & -\frac{5}{18} & 0 & 0 & 0 & 0 & 0 & -148050 & 0 \\
 7&0 & -\frac{29}{729} & 0 & 0 & 0 & 0 & 0 & \frac{989065}{324} & 0 & 0 & 0 \\
 8&0 & 0 & 0 & 0 & 0 & -\frac{8111}{243} & 0 & 0 & 0 & 0 & 0 \\
 9&0 & 0 & 0 & -\frac{1}{2} & 0 & 0 & 0 & 0 & 0 & \frac{4773195}{2} & 0 \\
 10&0 & -\frac{6607}{19683} & 0 & 0 & 0 & 0 & 0 & -\frac{81616115}{4374} & 0 & 0 & 0\\
&&&&&&&&&&&\\
\hline
\end{array}
\end{equation}



\section{Integrability of the Gauss-Manin system for more moduli}

The procedure outlined in the previous section for finding the right flat coordinates in moduli space generalizes to any number of closed and open moduli in the following way.  After finding the Picard-Fuchs operators we translate these into a matrix differential equation of first order for the vector
\begin{equation}
\begin{array}{cccccc}
\vec{\Omega}(z_.) &=&
(\underline{\Omega}(z_.)&
 \partial_{z_.} \underline{\Omega}(z_.) &
 \partial_{z_.}^2 \underline{\Omega}(z_.)&
\partial_{z_.}^3\underline{\Omega}(z_.)) \\
 \\
\# \, \textrm{entries}&= & d_0& d_1&d_2 &d_3
\\
 \\
&=&h^{3,0}(Z^*) &h^{2,1}(Z^*)+h^{2,0}(\cxH) &h^{1,2}(Z^*)+h^{1,1}(\cxH)&h^{0,3}(Z^*)+h^{0,2}(\cxH),
\end{array}
\end{equation}
where $d_p$ denotes the dimension of the space $F^{3-p} H^3$ of the Hodge filtration of the relative cohomology and corresponds to the number of charge $p$ states of the B-model in the subring of the chiral ring spanned by the marginal operators. The powers of the derivatives correspond to multi-derivatives in both open and closed moduli directions. The multi-derivatives of degree $2,3$ are linearly dependent. Linear independent combinations of these are picked to match the total number of independent elements given by the filtration. Now taking a derivative of $\vec{\Omega}(z_.)$ can be expressed as linear combinations of the entries of the vector by using the Picard-Fuchs equations and further operators obtained by linearly combining these. This linear dependence is expressed in the equation
\begin{equation}
\partial_{z_{.}} \vec{\underline{\Omega}}(z_{.})= M_{z_.} \vec{\underline{\Omega}}(z_{.}) \, .
\end{equation}
The connection matrices $M_{z_.}$ can be brought to the desired upper triangular form by  making an ansatz
\begin{equation}
\vec{\underline{\Omega}}'(t)=
\begin{array}{cccc}
(\underline{\Omega'}(t) &
\partial_{t_.} \underline{\Omega'}(t) &
C_{*}^{1}(t_.)^{-1}\partial_{t_.}^2 \underline{\Omega'}(t_.)&
C_{*}^{2}(t_.)^{-1}\partial_{*} (C_*^{1}(t_.)^{-1}\partial_{t_.}^2 \underline{\Omega'}(t_.) )\,,
\end{array}
\end{equation}
where $*$ denotes some fixed choice of closed string modulus for which the matrix $C^{(p)}_{*}$ is invertible and where
\begin{equation}
 \underline{\Omega}'(t_.) =\frac{\underline{\Omega}(z_.(t_.))}{\omega_0(t_.)}\, .
\end{equation}
The normalization factor $\omega_0(t_.)$ and the $d_q \times d_{q+1}$ matrices $C^{q}(t_.)$ are a priori unknown ans\"atze that have to be determined. In terms of these quantities and the mirror map ansatz $z(t)$, the matrix $A(t_.)$ relating  $\vec{\underline{\Omega}}(z_.)$ and $\vec{\underline{\Omega}}'(t_.)$ can be computed
$$ \vec{\underline{\Omega}}'(t_.)=A(t_.) \vec{\underline{\Omega(z_.)}} \, ,$$
and the new connection matrices $M_{t_.}$ can be computed out of $M_{z_.}$ and $A$:
\begin{equation}
M_{t_.} = \partial_{t_.} A \cdot A^{-1} + \sum_{z_.} \frac{\partial z_.}{\partial t_.} A \cdot M_{z_.} A^{-1}\,.
\end{equation}
Again the entries of the matrices $M_{t_.}$ which are not upper triangular give differential equations for the ans\"atze, which were made to rephrase the problem, and can be set to zero by solving the differential equations. The form of the variation of mixed Hodge structure now becomes
\begin{equation}
(\partial_{t_.} - M_{t_.}) \vec{\Omega}(t_.)=0\,,
\end{equation}
with
\begin{equation}
M_{t_.}= \left(\begin{array}{cccc}
0&C^{0}_{t_.} &0&0\\
0&0&C^{1}_{t_.}&0\\
0&0&0&C^{2}_{t_.}\\
0&0&0&0
\end{array}
\right) \, ,
\end{equation}
 where $C^{0}_{t_.}$ is a $1 \times d_1$ matrix having an entry $1$ corresponding to the position of the modulus $t_.$ under consideration. The $d_1 \times d_2$
matrix $C^{1}_{t_.}$ contains linear combinations of the generalization of the Yukawa couplings to the open string sector giving the OPE coefficient of the CFT pairing of charge 1 elements. The $d_2\times d_3$ matrix $C^2_{t_.}$ contains linear combinations of the generalization of the topological metric of the closed string sector which gives the pairing of charge 1 and charge 2 elements. In this case however it is moduli dependent.

In the following we will examine the structure of the entries of the matrices $C^{1}_{t_.}$ and $C^{2}_{t_.}$ imposed by the integrability and flatness requirements of the Gauss-Manin connection. It was shown in ref.\cite{Alim:2009bx} that the flatness and integrability of the Gauss-Manin connection imposes certain relations among the periods of the open-closed system. For the simplest example, \textit{i.e.}, one closed- and one open-string moduli, the K3-structure of the open-closed subsystem is sufficient to guarantee the integrability requirement \cite{Alim:2009bx}.  For cases with more moduli, one crucial task is to study the integrability structure and find the necessary relations among the relative periods, which are imposed by the integrability requirement. In this section, we consider two cases with more moduli and study the integrability structure. In the first case, we analyze an open-closed system with two closed- and one open-string moduli. The new aspect as compared to the case studied in \cite{Alim:2009bx} is that the subsystem may have a larger number of periods. To be concrete, we will focus on an example with two closed- and one open-string moduli, for which the associated open-closed subsystem has four linearly independent periods.

In the second case, we consider an open-closed system with one closed- and two open-string moduli. This example is particularly interesting, because the open-closed subsystem does not exhibit a K3-structure.  Therefore, it is different from the original example studied in ref.\cite{Alim:2009bx}.

In both cases, we first analyze the integrability structure by studying the Gauss-Manin connections in terms of appropriate flat coordinates in full generality. Explicit examples will be examined in section 6, as we address other interesting features of those examples. Since the first case is computationally similar to the example in \cite{Alim:2009bx}, we will be brief for the first part and only stress the differences. The reader can find more details in derivation of the second case as well as in ref.\cite{Alim:2009bx}.


\subsection{Two closed- and one open-string moduli}

In this section, we analyze the integrability of the Gauss-Manin system for the case of two closed- and one open-string moduli. We focus on a general case for which the open-closed subsystem exhibits a K3 structure. If the K3 subsystem has one algebraic modulus, then we have three relative periods for the subsystem and the situation is identical with the original example in ref.\cite{Alim:2009bx}. Examples of this type were studied in \cite{Jockers:2009mn,Alim:2010za}. However, depending on the choice of the divisor, the K3 subsystem may have two algebraic moduli (for explicit examples see refs.\cite{Jockers:2009mn,Alim:2010za}). In this case, one has four relative periods for the subsystem. We will now focus on this types and study integrability for them. As in ref.\cite{Alim:2009bx}, we compute the period matrix and bring it in the upper triangular form. This is the form that period matrix and connection matrices will take if we rewrite them in terms of the flat coordinates.

The periods associated with the relative holomorphic three-form of the open-closed system has the following form
\begin{eqnarray}\label{Omperiod}
(\underline{\Omega}:\, 1\qquad t_{1}\qquad t_{2}\qquad \hat{t}\qquad F_{t_{1}}\qquad F_{t_{2}}\quad W^{(1)}\quad W^{(2)}\quad -F_{0}\qquad T)\ ,
\end{eqnarray}
where $t_{1}$, $t_{2}$, and $\hat{t}$ are the closed- and open-string flat coordinates respectively. In (\ref{Omperiod}), $F$ is the prepotential of the closed sector, and therefore its derivatives $F_{t_1}$ and $F_{t_2}$ and the function $F_0$ depend on the closed string moduli $t_1$ and $t_2$, only. The remaining relative periods $W^{(1)}$, $W^{(2)}$, and $T$ of the open-closed system are functions of all moduli $t_1$, $t_2$ and $\hat t$. Since the open-closed subsystem is associated to a K3 surface, there is one relation among the subsystem periods, which is obtained from the full system periods by differentiating with respect to $\hat{t}$
\begin{eqnarray}\label{dtopT}
T_{\hat{t}}=\frac{1}{2}\,\overrightarrow{W}_{\hat{t}}\cdot\eta\cdot
\overrightarrow{W}_{\hat{t}}\ ,
\end{eqnarray}
where $\overrightarrow{W}=(W^{(1)}\quad W^{(2)})$ and $\eta$ is the intersection form of the middle cohomology group of the K3 subsystem. Similar to the original case in \cite{Alim:2009bx}, one can integrate this relation into a functional form for the top period $T$
\begin{eqnarray}\label{topT}
T(t_{1},t_{2},\hat{t})=\frac{1}{2}\int_{0}^{1} d\sigma\,\,\, \hat{t}\left(\overrightarrow{W}_{\hat{t}}(t_{1},t_{2},\sigma\hat{t})\cdot\eta\cdot
\overrightarrow{W}_{\hat{t}}(t_{1},t_{2},\sigma\hat{t})\right)+f(t_{1},t_{2})\ ,
\end{eqnarray}
where $f(t_{1},t_{2})$ is a function of the two closed-string moduli, which appears as the constant of integration. In order to obtain the connection matrices in flat coordinates, we need to bring the matrices into upper triangular form. To do this, we need the second derivatives of the top periods. They can be deduced from (\ref{dtopT})
\begin{eqnarray}\label{derT}
T_{\hat{t}\,\hat{t}}=\overrightarrow{W}_{\hat{t}}\cdot\eta\cdot\overrightarrow{W}_
{\hat{t}\,\hat{t}}\qquad,\qquad  T_{t_{i}\,\hat{t}}=\overrightarrow{W}_{\hat{t}}\cdot\eta\cdot\overrightarrow{W}_
{t_{i}\,\hat{t}}\ ,\quad i\in\{1,2\}\ .
\end{eqnarray}
After a some steps of algebra,\footnote{This computation is similar to what is done in ref.\cite{Alim:2009bx} and we only mention the results here. Further details will be revealed for the next example, which also exhibits new features.} the connection matrices are found
\begin{eqnarray}\label{Mtone}
M_{t_{1}}(t_{1},t_{2},\hat{t})=
\left(
  \begin{array}{cccccccccc}
    0 & 1 & 0 & 0 & 0 & 0 & 0 & 0 & 0 & 0 \\
    0 & 0 & 0 & 0 & F_{t_{1}t_{1}t_{1}} & F_{t_{1}t_{1}t_{2}} & W^{(1)}_{t_{1}t_{1}} & W^{(2)}_{t_{1}t_{1}} & 0 & 0 \\
    0 & 0 & 0 & 0 & F_{t_{1}t_{1}t_{2}} & F_{t_{1}t_{2}t_{2}} & W^{(1)}_{t_{1}t_{2}} & W^{(2)}_{t_{1}t_{2}} & 0 & \rho \\
    0 & 0 & 0 & 0 & 0 & 0 & W^{(1)}_{t_{1}\hat{t}} & W^{(2)}_{t_{1}\hat{t}} & 0 & 0 \\
    0 & 0 & 0 & 0 & 0 & 0 & 0 & 0 & 1 & \partial_{t_{1}}\mu_{1} \\
    0 & 0 & 0 & 0 & 0 & 0 & 0 & 0 & 0 & \partial_{t_{1}}\mu_{2} \\
    0 & 0 & 0 & 0 & 0 & 0 & 0 & 0 & 0 & W^{(2)}_{t_{1}\hat{t}} \\
    0 & 0 & 0 & 0 & 0 & 0 & 0 & 0 & 0 & W^{(1)}_{t_{1}\hat{t}} \\
    0 & 0 & 0 & 0 & 0 & 0 & 0 & 0 & 0 & 0 \\
    0 & 0 & 0 & 0 & 0 & 0 & 0 & 0 & 0 & 0 \\
  \end{array}
\right)\ ,
\end{eqnarray}
for $M_{t_{2}}$, we find
\begin{eqnarray}\label{Mttwo}
M_{t_{2}}(t_{1},t_{2},\hat{t})=
\left(
  \begin{array}{cccccccccc}
    0 & 0 & 1 & 0 & 0 & 0 & 0 & 0 & 0 & 0 \\
    0 & 0 & 0 & 0 & F_{t_{1}t_{1}t_{2}} & F_{t_{1}t_{2}t_{2}} & W^{(1)}_{t_{1}t_{2}} & W^{(2)}_{t_{1}t_{2}} & 0 & \rho \\
    0 & 0 & 0 & 0 & F_{t_{1}t_{2}t_{2}} & F_{t_{2}t_{2}t_{2}} & W^{(1)}_{t_{2}t_{2}} & W^{(2)}_{t_{2}t_{2}} & 0 & 0 \\
    0 & 0 & 0 & 0 & 0 & 0 & W^{(1)}_{t_{2}\hat{t}} & W^{(2)}_{t_{2}\hat{t}} & 0 & 0 \\
    0 & 0 & 0 & 0 & 0 & 0 & 0 & 0 & 0 & \partial_{t_{2}}\mu_{1} \\
    0 & 0 & 0 & 0 & 0 & 0 & 0 & 0 & 1 & \partial_{t_{2}}\mu_{2} \\
    0 & 0 & 0 & 0 & 0 & 0 & 0 & 0 & 0 & W^{(2)}_{t_{2}\hat{t}} \\
    0 & 0 & 0 & 0 & 0 & 0 & 0 & 0 & 0 & W^{(1)}_{t_{2}\hat{t}} \\
    0 & 0 & 0 & 0 & 0 & 0 & 0 & 0 & 0 & 0 \\
    0 & 0 & 0 & 0 & 0 & 0 & 0 & 0 & 0 & 0 \\
  \end{array}
\right)\ ,
\end{eqnarray}
and finally for $M_{\hat{t}}$, we find
\begin{eqnarray}\label{Mttwoii}
M_{\hat{t}}(t_{1},t_{2},\hat{t})=
\left(
  \begin{array}{cccccccccc}
    0 & 0 & 0 & 1 & 0 & 0 & 0 & 0 & 0 & 0 \\
    0 & 0 & 0 & 0 & 0 & 0 & W^{(1)}_{t_{1}\hat{t}} & W^{(2)}_{t_{1}\hat{t}} & 0 & 0 \\
    0 & 0 & 0 & 0 & 0 & 0 & W^{(1)}_{t_{2}\hat{t}} & W^{(2)}_{t_{2}\hat{t}} & 0 & 0 \\
    0 & 0 & 0 & 0 & 0 & 0 & W^{(1)}_{\hat{t}\hat{t}} & W^{(2)}_{\hat{t}\hat{t}} & 0 & 0 \\
    0 & 0 & 0 & 0 & 0 & 0 & 0 & 0 & 0 & \partial_{\hat{t}}\mu_{1} \\
    0 & 0 & 0 & 0 & 0 & 0 & 0 & 0 & 0 & \partial_{\hat{t}}\mu_{2} \\
    0 & 0 & 0 & 0 & 0 & 0 & 0 & 0 & 0 & W^{(2)}_{\hat{t}\hat{t}} \\
    0 & 0 & 0 & 0 & 0 & 0 & 0 & 0 & 0 & W^{(1)}_{\hat{t}\hat{t}} \\
    0 & 0 & 0 & 0 & 0 & 0 & 0 & 0 & 0 & 0 \\
    0 & 0 & 0 & 0 & 0 & 0 & 0 & 0 & 0 & 0 \\
  \end{array}
\right)\ .
\end{eqnarray}
The functions $\mu_{1}$, and $\mu_{2}$ which appear in above connection matrices are defined as
\begin{eqnarray}
\mu_{1}&=&+\frac{F_{t_{2}t_{2}t_{2}}}{C}\Big(T_{t_{1}t_{1}}-\overrightarrow{W}_{\hat{t}}
\cdot\eta\cdot\overrightarrow{W}_{t_{1}t_{1}}\Big)-
\frac{F_{t_{1}t_{1}t_{2}}}{C}\Big(T_{t_{2}t_{2}}-\overrightarrow{W}_{\hat{t}}
\cdot\eta\cdot\overrightarrow{W}_{t_{2}t_{2}}\Big)\ ,\label{muone}\\
\mu_{2}&=&-\frac{F_{t_{1}t_{2}t_{2}}}{C}\Big(T_{t_{1}t_{1}}-\overrightarrow{W}_{\hat{t}}
\cdot\eta\cdot\overrightarrow{W}_{t_{1}t_{1}}\Big)+
\frac{F_{t_{1}t_{1}t_{1}}}{C}\Big(T_{t_{2}t_{2}}-\overrightarrow{W}_{\hat{t}}
\cdot\eta\cdot\overrightarrow{W}_{t_{2}t_{2}}\Big)\ ,\label{mtwo}
\end{eqnarray}
where $C$ is the determinant of the Yukawa couplings
\begin{eqnarray}\label{yukadet}
C=\left|\begin{array}{cc}
          F_{t_{1}t_{1}t_{1}} & F_{t_{1}t_{1}t_{2}} \\
          F_{t_{1}t_{2}t_{2}} & F_{t_{2}t_{2}t_{2}}
        \end{array}
\right|\ .
\end{eqnarray}
In contrast to the case of one closed-string and one open-string modulus, in the connection matrix there appears another off-diagonal element encoded in function $\rho$
\begin{eqnarray}\label{rhofunc}
\rho&=&T_{t_{1}t_{2}}-\mu_{1}\,F_{t_{1}t_{1}t_{2}}-\mu_{2}\,F_{t_{1}t_{2}t_{2}}-
\overrightarrow{W}_{\hat{t}}\cdot\eta\cdot\overrightarrow{W}_{t_{1}t_{2}}\ .
\end{eqnarray}
If one now computes the curvature of the above Gauss-Manin connection, using \eqref{derT} one finds the flatness condition
\begin{equation}\label{flat}
  [\nabla_{t_{1}},\nabla_{t_{2}}]=0\ , \quad
  [\nabla_{\hat{t}},\nabla_{t_{1}}]=0\ ,\quad
  [\nabla_{\hat{t}},\nabla_{t_{2}}]=0\ .
\end{equation}
In summary, the integrability requirement for two closed- and one open-string moduli case leads to the same constraints as for the case of one closed- and one open-string moduli. Nonetheless, the Gauss-Manin connection matrices have new non-vanishing off-diagonal elements.

\subsection{One closed- and two open-string moduli}\label{onetwomod}

So far, we have considered examples in which the open-closed system has one open-string modulus. It is also interesting to analyze examples with more than one open-string modulus as well. This case is particularly interesting because the Hodge variation of the subsystem is not equivalent to that on a K3 manifold anymore. The easiest way to realize this is that the subsystem, as a complete intersection manifold, now has more than one independent holomorphic two-form, each associated with one open-string modulus.\footnote{In the diagram of Hodge variation, the derivative of the holomorphic three-form with respect to each open modulus descends to a holomorphic two-form of the subsystem.} Therefore, it is important to investigate what relations are imposed in this case by the integrability requirement. For concreteness, we consider an open-closed system consisting one closed- and two open-string moduli. We examine an explicit example of this type in the next section. As this example reveals a different structure, we now analyze the relevant Gauss-Manin system in detail.

To start, we first construct the period matrix of the open-closed system in terms of the flat coordinates of the system and bring it into an upper triangular form. At the level of zero charge, we have the following ansatz for the period vector
\begin{eqnarray}\label{zerocharge}
q=0\,:\qquad (\Omega:\ 1\quad t\quad \hat{t}_{1}\quad \hat{t}_{2}\quad F_{t}\quad W^{(1)}\quad W^{(2)}\quad -F_{0}\quad T^{(1)}\quad T^{(2)})\ ,
\end{eqnarray}
where $t$ is the bulk flat coordinate whereas $\hat{t}_{1}$ and $\hat{t}_{2}$ are the two open flat coordinates. Defining $\phi_{1}\equiv\partial_{t}\Omega$\ ,\ $\hat{\phi}_{1}\equiv\partial_{\hat{t}_{1}}\Omega$\ ,\ $\tilde{\phi}_{1}\equiv\partial_{\hat{t}_{2}}\Omega$\ , at level $q=1$, the periods read
\begin{eqnarray}\label{chargeone}
q=1:\quad\left(
               \begin{array}{ccccccccccc}
                 \phi_{1}: & 0 & 1 & 0& 0& F_{tt}& W^{(1)}_{t}& W^{(2)}_{t}& -F_{0t}& T^{(1)}_{t}& T^{(2)}_{t}\cr
\hat{\phi}_{1}: & 0 & 0 & 1& 0& 0& W^{(1)}_{\hat{t}_{1}}& W^{(2)}_{\hat{t}_{1}}& 0& T^{(1)}_{\hat{t}_{1}}& T^{(2)}_{\hat{t}_{1}}\cr
\tilde{\phi}_{1}: & 0 & 0 & 0& 1& 0& W^{(1)}_{\hat{t}_{2}}& W^{(2)}_{\hat{t}_{2}}& 0& T^{(1)}_{\hat{t}_{2}}& T^{(2)}_{\hat{t}_{2}}
               \end{array}
             \right)
\ ,
\end{eqnarray}
and for the derivatives of these periods, we find
\begin{eqnarray}\label{chargeoneder}
\left(
  \begin{array}{ccccccccccc}
    \partial_{t}\phi_{1}: & 0 & 0 & 0& 0& C& W^{(1)}_{tt}& W^{(2)}_{tt}& t\,C & T^{(1)}_{tt}& T^{(2)}_{tt}\cr
\partial_{t}\hat{\phi}_{1}: & 0 & 0 & 0& 0& 0& W^{(1)}_{t \hat{t}_{1}}& W^{(2)}_{t \hat{t}_{1}}& 0& T^{(1)}_{t \hat{t}_{1}}& T^{(2)}_{t \hat{t}_{1}}\cr
\partial_{t}\tilde{\phi}_{1}: & 0 & 0 & 0& 0& 0& W^{(1)}_{t \hat{t}_{2}}& W^{(2)}_{t \hat{t}_{2}}& 0& T^{(1)}_{t \hat{t}_{2}}& T^{(2)}_{t \hat{t}_{2}}
  \end{array}
\right)
\ ,
\end{eqnarray}
where $C=F_{ttt}$ as usual and we used $-F_{0tt}=C\,t$. Straightforwardly, for the level $q=2$ charge we obtain
\begin{eqnarray}\label{chargetwo}
q=2:\quad\left(
           \begin{array}{ccccccccccc}
             \phi_{2}: & 0 & 0 & 0& 0& 1& 0& 0& t& \mu& \nu\cr
\hat{\phi}_{2}: & 0 & 0 & 0& 0& 0& 1& 0& 0& \hat{\mu}& \hat{\nu}\cr
\tilde{\phi}_{2}: & 0 & 0 & 0& 0& 0& 0& 1& 0& \tilde{\mu}& \tilde{\nu}
           \end{array}
         \right)
\ ,
\end{eqnarray}
where the fields at $q=2$ are defined as:
\begin{eqnarray}\label{fieldchtwo}
\phi_{2}&=&{1\over C\, D}(D\,\partial_{t}\phi_{1}-D_{2}\partial_{t}\hat{\phi}_{1}+D_{1}\partial_{t}\tilde{\phi}_{1})\ ,\cr
\hat{\phi}_{2}&=&{1\over D}(W^{(2)}_{t \hat{t}_{2}}\partial_{t}\hat{\phi}_{1}-W^{(2)}_{t \hat{t}_{1}}\partial_{t}\tilde{\phi}_{1})\ ,\cr
\tilde{\phi}_{2}&=&{1\over D}(-W^{(1)}_{t \hat{t}_{2}}\partial_{t}\hat{\phi}_{1}+W^{(1)}_{t \hat{t}_{1}}\partial_{t}\tilde{\phi}_{1})\ ,
\end{eqnarray}
in which $D$, $D_{1}$, and $D_{2}$ are defined as follows:
\begin{eqnarray}\label{defD}
D\equiv\left|{\partial(W^{(1)}_{t},W^{(2)}_{t})\over \partial(\hat{t}_{1},\hat{t}_{2})}\right|\quad,\quad D_{1}\equiv\left|{\partial(W^{(1)}_{t},W^{(2)}_{t})\over \partial(t,\hat{t}_{1})}\right|\quad,\quad D_{2}\equiv\left|{\partial(W^{(1)}_{t},W^{(2)}_{t})\over \partial(t,\hat{t}_{2})}\right|\ .
\end{eqnarray}
We also need to define the ``potentials'' $(\mu,\nu)$, $(\hat{\mu},\hat{\nu})$, and $(\tilde{\mu},\tilde{\nu})$ appearing in (\ref{chargetwo}). The first pair is defined as:
\begin{eqnarray}\label{defmunu}
\mu={1\over C\,D}\left|
\begin{array}{ccc}
  W^{(1)}_{t \hat{t}_{1}} & W^{(1)}_{t \hat{t}_{2}} & W^{(1)}_{t t} \cr W^{(2)}_{t \hat{t}_{1}} & W^{(2)}_{t \hat{t}_{2}} & W^{(2)}_{t t} \cr T^{(1)}_{t \hat{t}_{1}} &
T^{(1)}_{t \hat{t}_{2}} & T^{(1)}_{t t} \end{array}\right|\quad,\quad
\nu={1\over C\,D}\left|
\begin{array}{ccc}
  W^{(1)}_{t \hat{t}_{1}} & W^{(1)}_{t \hat{t}_{2}} & W^{(1)}_{t t} \cr W^{(2)}_{t \hat{t}_{1}} & W^{(2)}_{t \hat{t}_{2}} & W^{(2)}_{t t} \cr T^{(2)}_{t \hat{t}_{1}} &
T^{(2)}_{t \hat{t}_{2}} & T^{(2)}_{t t}
\end{array}
\right|\ ,
\end{eqnarray}
and the remaining potentials are given by the following expressions
\begin{eqnarray}\label{defhatmunu}
\hat{\mu}&=&-{1\over D}\left|{\partial(W^{(2)}_{t},T^{(1)}_{t})\over\partial(\hat{t}_{1},\hat{t}_{2})}\right|\quad,\quad \hat{\nu}=-{1\over D}\left|{\partial(W^{(2)}_{t},T^{(2)}_{t})\over\partial(\hat{t}_{1},\hat{t}_{2})}\right|\ ,\cr
\tilde{\mu}&=&{1\over D}\left|{\partial(W^{(1)}_{t},T^{(1)}_{t})\over\partial(\hat{t}_{1},\hat{t}_{2})}\right|\quad,\quad \tilde{\nu}={1\over D}\left|{\partial(W^{(1)}_{t},T^{(2)}_{t})\over\partial(\hat{t}_{1},\hat{t}_{2})}\right|\ .
\end{eqnarray}
Finally, for the last block of the period matrix, level $q=3$, we obtain
\begin{eqnarray}\label{chargethree}
q=3:\quad\left(
           \begin{array}{ccccccccccc}
            \phi_{3}: & 0 & 0 & 0& 0& 0& 0& 0& 1& 0& 0\cr
\hat{\phi}_{3}: & 0 & 0 & 0& 0& 0& 0& 0& 0& 1& 0\cr
\tilde{\phi}_{3}: & 0 & 0 & 0& 0& 0& 0& 0& 0& 0& 1
           \end{array}
         \right)
\ ,
\end{eqnarray}
where the chiral fields $\phi_{3}$, $\hat{\phi}_{3}$, and $\tilde{\phi}_{3}$ are defined as
\begin{eqnarray}\label{fieldchthree}
\left(
  \begin{array}{c}
    \phi_{3}\cr \hat{\phi}_{3}\cr \tilde{\phi}_{3}
  \end{array}
\right)
=\left(
   \begin{array}{ccc}
     1& \partial_{t}\mu& \partial_{t}\nu\cr 0& \partial_{t}\hat{\mu}& \partial_{t}\hat{\nu}\cr 0& \partial_{t}\tilde{\mu}& \partial_{t}\tilde{\nu}
   \end{array}
 \right)^{-1}\left(
               \begin{array}{c}
                \partial_{t}\phi_{2}\cr \partial_{t}\hat{\phi}_{2}\cr \partial_{t}\tilde{\phi}_{2}
               \end{array}
             \right)
\ .
\end{eqnarray}
Therefore, the full period matrix for this system in terms of the flat coordinates reads
\begin{eqnarray}\label{periodmat}
\Pi(t,\hat{t}_{1},\hat{t}_{2})=\left(
                                 \begin{array}{cccccccccc}
                                   1 & t & \hat{t}_{1} & \hat{t}_{2} & F_{t} & W^{(1)} & W^{(2)} & -F_{0} & T^{(1)} & T^{(2)} \cr 0 & 1 & 0& 0& F_{tt}& W^{(1)}_{t}& W^{(2)}_{t}& -F_{0t}& T^{(1)}_{t}& T^{(2)}_{t}\cr 0 & 0 & 1& 0& 0& W^{(1)}_{\hat{t}_{1}}& W^{(2)}_{\hat{t}_{1}}& 0& T^{(1)}_{\hat{t}_{1}}& T^{(2)}_{\hat{t}_{1}}\cr 0 & 0 & 0& 1& 0& W^{(1)}_{\hat{t}_{2}}& W^{(2)}_{\hat{t}_{2}}& 0& T^{(1)}_{\hat{t}_{2}}& T^{(2)}_{\hat{t}_{2}}\cr 0 & 0 & 0& 0& 1& 0& 0& t& \mu& \nu\cr 0 & 0 & 0& 0& 0& 1& 0& 0& \hat{\mu}& \hat{\nu}\cr 0 & 0 & 0& 0& 0& 0& 1& 0& \tilde{\mu}& \tilde{\nu}\cr 0 & 0 & 0& 0& 0& 0& 0& 1& 0& 0\cr 0 & 0 & 0& 0& 0& 0& 0& 0& 1& 0\cr 0 & 0 & 0& 0& 0& 0& 0& 0& 0& 1
                                 \end{array}
                               \right)
\ .
\end{eqnarray}
Using $M_{a}=(\partial_{a}\Pi)\cdot\Pi^{-1}$, we can compute the Gauss-Manin connection matrices of this system in full generality. We have three connection matrices and they are  found to be
\begin{eqnarray}\label{Mt}
M_{t}(t,\hat{t}_{1},\hat{t}_{2})=
\left(
  \begin{array}{cccccccccc}
    0 & 1 & 0 & 0 & 0 & 0 & 0 & 0 & 0 & 0 \cr
  0 & 0 & 0 & 0 & C & W^{(1)}_{t t} & W^{(2)}_{t t} & 0 & 0 & 0 \cr
  0 & 0 & 0 & 0 & 0 & W^{(1)}_{t \hat{t}_{1}} & W^{(2)}_{t \hat{t}_{1}} & 0 & 0 & 0 \cr
  0 & 0 & 0 & 0 & 0 & W^{(1)}_{t \hat{t}_{2}} & W^{(2)}_{t \hat{t}_{2}} & 0 & 0 & 0 \cr
  0 & 0 & 0 & 0 & 0 & 0 & 0 & 1 & \partial_{t}\mu & \partial_{t}\nu \cr
  0 & 0 & 0 & 0 & 0 & 0 & 0 & 0 & \partial_{t}\hat{\mu} & \partial_{t}\hat{\nu} \cr
  0 & 0 & 0 & 0 & 0 & 0 & 0 & 0 & \partial_{t}\tilde{\mu} & \partial_{t}\tilde{\nu} \cr
  0 & 0 & 0 & 0 & 0 & 0 & 0 & 0 & 0 & 0 \cr
  0 & 0 & 0 & 0 & 0 & 0 & 0 & 0 & 0 & 0 \cr
  0 & 0 & 0 & 0 & 0 & 0 & 0 & 0 & 0 & 0
  \end{array}
\right)
  \ ,
\end{eqnarray}
for $M_{\hat{t}_{1}}$:
\begin{eqnarray}\label{Mthatone}
M_{\hat{t}_{1}}(t,\hat{t}_{1},\hat{t}_{2})=
\left(
  \begin{array}{cccccccccc}
    0 & 0 & 1 & 0 & 0 & 0 & 0 & 0 & 0 & 0 \cr
  0 & 0 & 0 & 0 & 0 & W^{(1)}_{t \hat{t}_{1}} & W^{(2)}_{t \hat{t}_{1}} & 0 & 0 & 0 \cr
  0 & 0 & 0 & 0 & 0 & W^{(1)}_{\hat{t}_{1} \hat{t}_{1}} & W^{(2)}_{\hat{t}_{1} \hat{t}_{1}} & 0 & A_{1} & A_{2} \cr
  0 & 0 & 0 & 0 & 0 & W^{(1)}_{\hat{t}_{1} \hat{t}_{2}} & W^{(2)}_{\hat{t}_{1} \hat{t}_{2}} & 0 & A_{3} & A_{4} \cr
  0 & 0 & 0 & 0 & 0 & 0 & 0 & 0 & \partial_{\hat{t}_{1}}\mu & \partial_{\hat{t}_{1}}\nu \cr
  0 & 0 & 0 & 0 & 0 & 0 & 0 & 0 & \partial_{\hat{t}_{1}}\hat{\mu} & \partial_{\hat{t}_{1}}\hat{\nu} \cr
  0 & 0 & 0 & 0 & 0 & 0 & 0 & 0 & \partial_{\hat{t}_{1}}\tilde{\mu} & \partial_{\hat{t}_{1}}\tilde{\nu} \cr
  0 & 0 & 0 & 0 & 0 & 0 & 0 & 0 & 0 & 0 \cr
  0 & 0 & 0 & 0 & 0 & 0 & 0 & 0 & 0 & 0 \cr
  0 & 0 & 0 & 0 & 0 & 0 & 0 & 0 & 0 & 0
  \end{array}
\right)
\ ,
\end{eqnarray}
and finally for $M_{\hat{t}_{2}}$:
\begin{eqnarray}\label{Mthattwo}
M_{\hat{t}_{2}}(t,\hat{t}_{1},\hat{t}_{2})=
\left(
  \begin{array}{cccccccccc}
    0 & 0 & 0 & 1 & 0 & 0 & 0 & 0 & 0 & 0 \cr
  0 & 0 & 0 & 0 & 0 & W^{(1)}_{t \hat{t}_{2}} & W^{(2)}_{t \hat{t}_{2}} & 0 & 0 & 0 \cr
  0 & 0 & 0 & 0 & 0 & W^{(1)}_{\hat{t}_{1} \hat{t}_{2}} & W^{(2)}_{\hat{t}_{1} \hat{t}_{2}} & 0 & A_{3} & A_{4} \cr
  0 & 0 & 0 & 0 & 0 & W^{(1)}_{\hat{t}_{2} \hat{t}_{2}} & W^{(2)}_{\hat{t}_{2} \hat{t}_{2}} & 0 & A_{5} & A_{6} \cr
  0 & 0 & 0 & 0 & 0 & 0 & 0 & 0 & \partial_{\hat{t}_{2}}\mu & \partial_{\hat{t}_{2}}\nu \cr
  0 & 0 & 0 & 0 & 0 & 0 & 0 & 0 & \partial_{\hat{t}_{2}}\hat{\mu} & \partial_{\hat{t}_{2}}\hat{\nu} \cr
  0 & 0 & 0 & 0 & 0 & 0 & 0 & 0 & \partial_{\hat{t}_{2}}\tilde{\mu} & \partial_{\hat{t}_{2}}\tilde{\nu} \cr
  0 & 0 & 0 & 0 & 0 & 0 & 0 & 0 & 0 & 0 \cr
  0 & 0 & 0 & 0 & 0 & 0 & 0 & 0 & 0 & 0 \cr
  0 & 0 & 0 & 0 & 0 & 0 & 0 & 0 & 0 & 0
  \end{array}
\right)
\ .
\end{eqnarray}
In the above expressions, the functions $A_{i}$'s are given by
\begin{eqnarray}\label{As}
A_{1}&=&T^{(1)}_{\hat{t}_{1} \hat{t}_{1}}-\hat{\mu}\,W^{(1)}_{\hat{t}_{1} \hat{t}_{1}}-\tilde{\mu}\,W^{(2)}_{\hat{t}_{1} \hat{t}_{1}}\quad,\quad  A_{2}=T^{(2)}_{\hat{t}_{1} \hat{t}_{1}}-\hat{\nu}\,W^{(1)}_{\hat{t}_{1} \hat{t}_{1}}-\tilde{\nu}\,W^{(2)}_{\hat{t}_{1} \hat{t}_{1}}\ ,\cr
A_{3}&=&T^{(1)}_{\hat{t}_{1} \hat{t}_{2}}-\hat{\mu}\,W^{(1)}_{\hat{t}_{1} \hat{t}_{2}}-\tilde{\mu}\,W^{(2)}_{\hat{t}_{1} \hat{t}_{2}}\quad,\quad  A_{4}=T^{(2)}_{\hat{t}_{1} \hat{t}_{2}}-\hat{\nu}\,W^{(1)}_{\hat{t}_{1} \hat{t}_{2}}-\tilde{\nu}\,W^{(2)}_{\hat{t}_{1} \hat{t}_{2}}\ ,\cr
A_{5}&=&T^{(1)}_{\hat{t}_{2} \hat{t}_{2}}-\hat{\mu}\,W^{(1)}_{\hat{t}_{2} \hat{t}_{2}}-\tilde{\mu}\,W^{(2)}_{\hat{t}_{2} \hat{t}_{2}}\quad,\quad  A_{6}=T^{(2)}_{\hat{t}_{2} \hat{t}_{2}}-\hat{\nu}\,W^{(1)}_{\hat{t}_{2} \hat{t}_{2}}-\tilde{\nu}\,W^{(2)}_{\hat{t}_{2} \hat{t}_{2}}\ .
\end{eqnarray}

\subsubsection{Integrability of the Gauss-Manin connection}

The connection matrices we have derived so far will not fulfill the integrability condition, unless there exist relations among the periods. We can now explicitly compute the curvatures associated with connection matrices (\ref{Mt}), (\ref{Mthatone}), and (\ref{Mthattwo}). There are three relations coming from the curvatures and if we require them to vanish, we get three different types of relations among the relative periods:
\begin{enumerate}
\item \textit{Relations among the first derivative of the periods:}

There are three relations among the first derivative of the periods with respect to the open flat coordinates. These relations are associated with the periods of the subsystem, in which case is a complex algebraic surface. Obviously, the subsystem in this case is not associated to a K3 surface and these relations are considered as the generalization of the simple relation among the K3 periods (schematically $\omega_{1}^{2}=\omega_{0}\omega_{2}$). To present the  relations, let us use a compact notation and introduce the vector $\overrightarrow{W}=(W^{(1)}\quad W^{(2)})$,  equipped with a constant metric $\eta$ acting on this subspace. The three relations are then given by
\begin{eqnarray}\label{relone}
  T^{(1)}_{\hat{t}_1}=\frac{1}{2}\,\overrightarrow{W}_{\hat t_1}\cdot\eta\cdot \overrightarrow{W}_{\hat t_1} \ , \
  T^{(2)}_{\hat{t}_2}=\frac{1}{2}\,\overrightarrow{W}_{\hat t_2}\cdot\eta\cdot \overrightarrow{W}_{\hat t_2} \ , \
  T^{(1)}_{\hat{t}_2}+T^{(2)}_{\hat{t}_1}=\overrightarrow{W}_{\hat t_1}\cdot\eta\cdot \overrightarrow{W}_{\hat t_2} \ ,
\end{eqnarray}
where we have chosen an overall normalization factor 1/2 for ease of notation and for later purposes. The explicit form of the matrix $\eta$ depends on properties of the analyzed open-closed system, and a chosen superpotential period singles out a particular brane configuration. For the concrete example of the next section, we will work the matrix $\eta$ explicitly. Note that, in contrast to the one open modulus case (described by K3 periods for the subsystem), the $T^{(i)}$ periods do not have a functional representation in terms of the superpotential periods.

\item  \textit{Relations for the second derivatives of the periods}

In addition to (\ref{relone}), we also have relations among the second derivatives of the periods. Some of these relations are simply obtained by differentiating (\ref{relone}), and therefore they are not independent relations. However, there are additional relations, which cannot be obtained by differentiation of the previous relations. These relations are given by
\begin{eqnarray}\label{reltwo}
T^{(i)}_{t \hat{t}_{j}}&=&\overrightarrow{W}_{\hat{t}_{i}}\cdot\eta\cdot\overrightarrow{W}_{t \hat{t}_{j}}\ ,\quad\qquad i,j\in\{1,2\}\cr
T^{(i)}_{\hat{t}_{j} \hat{t}_{k}}&=&\overrightarrow{W}_{\hat{t}_{i}}\cdot\eta\cdot\overrightarrow{W}_{\hat{t}_{j} \hat{t}_{k}}\ ,\qquad i,j,k\in\{1,2\}
\end{eqnarray}
and do not depend on $T^{(1)}_{t t}$ and $T^{(2)}_{t t}$.

\item \textit{Relations among the derivatives of superpotential periods}

Integrability condition also imposes another type of relations which only involve the superpotential periods. They are given by
\begin{eqnarray}\label{relthree}
\overrightarrow{W}_{t \hat{t}_{1}}\cdot\eta\cdot\overrightarrow{W}_{\hat{t}_{1}\hat{t}_{2}}-
\overrightarrow{W}_{\hat{t}_{1}\hat{t}_{1}}\cdot\eta\cdot\overrightarrow{W}_{t \hat{t}_{2}}&=&\ 0\ ,\cr
\overrightarrow{W}_{t \hat{t}_{2}}\cdot\eta\cdot\overrightarrow{W}_{\hat{t}_{1}\hat{t}_{2}}-
\overrightarrow{W}_{t \hat{t}_{1}}\cdot\eta\cdot\overrightarrow{W}_{\hat{t}_{2} \hat{t}_{2}}&=&\ 0\ ,\cr
\overrightarrow{W}_{\hat{t}_{1} \hat{t}_{1}}\cdot\eta\cdot\overrightarrow{W}_{\hat{t}_{2}\hat{t}_{2}}-
\overrightarrow{W}_{\hat{t}_{1}\hat{t}_{2}}\cdot\eta\cdot\overrightarrow{W}_{\hat{t}_{1} \hat{t}_{2}}&=&\ 0\ .
\end{eqnarray}
Again note that $W^{(1)}_{t t}$ and $W^{(1)}_{t t}$ do not enter \eqref{relthree}.
\end{enumerate}
Having (\ref{relone}), (\ref{reltwo}), and (\ref{relthree}) which guarantee the integrability of the Gauss-Manin connection, we can simplify the connection matrices. First of all, it is easy to see that by (\ref{reltwo}), all $A_{i}$'s vanish. For instance, let us consider $A_{1}$. From (\ref{As}), one can show that
\begin{eqnarray}\label{Aone}
A_{1}={1\over D}\left|\begin{array}{ccc}
                        W^{(1)}_{t \hat{t}_{1}} & W^{(1)}_{t \hat{t}_{2}} & W^{(1)}_{\hat{t}_{1} \hat{t}_{1}} \cr W^{(2)}_{t \hat{t}_{1}} & W^{(2)}_{t \hat{t}_{2}} & W^{(2)}_{\hat{t}_{1} \hat{t}_{1}} \cr T^{(1)}_{t \hat{t}_{1}} & T^{(1)}_{t \hat{t}_{2}} & T^{(1)}_{\hat{t}_{1} \hat{t}_{1}}
                      \end{array}
\right|\ ,
\end{eqnarray}
and if one now substitutes the last row of the determinant by (\ref{reltwo}), one immediately finds that $A_{1}=0$. Also, using the same relation, we easily find
\begin{eqnarray}\label{musnus}
\hat{\mu}=W^{(2)}_{\hat{t}_{1}}\ ,\ \hat{\nu}=W^{(2)}_{\hat{t}_{2}}\ ,\ \tilde{\mu}=W^{(1)}_{\hat{t}_{1}}\ ,\ \tilde{\nu}=W^{(1)}_{\hat{t}_{2}}\ .
\end{eqnarray}
Using (\ref{reltwo}) for (\ref{defmunu}), we can also simplify $\mu$ and $\nu$, which appear in the period matrix:
\begin{eqnarray}\label{munurev}
\mu={1\over C}(T^{(1)}_{t t}-\overrightarrow{W}_{\hat{t}_{1}}\cdot\eta\cdot\overrightarrow{W}_{t t})\qquad,\qquad \nu={1\over C}(T^{(2)}_{t t}-\overrightarrow{W}_{\hat{t}_{2}}\cdot\eta\cdot\overrightarrow{W}_{t t})\ .
\end{eqnarray}
As $\partial_{\hat t_2}\mu\equiv\partial_{\hat t_1}\nu$, $\mu$ and $\nu$ are integrable to a generating function $G$
\begin{eqnarray}\label{GPot}
G(t,\hat t_1,\hat t_2)\,=\, \int_0^1 d\sigma \left(\hat t_1\, \mu(t,\sigma \hat t_1,\sigma \hat t_2)+\hat t_2\, \nu(t,\sigma \hat t_1,\sigma \hat t_2) \right) +g(t)\ ,
\end{eqnarray}
namely
\begin{eqnarray}\label{GRel}
G_{\hat t_1} \,=\, \mu \ , \quad G_{\hat t_2}\,=\,\nu \ .
\end{eqnarray}
We notice that $g(t)$ is an arbitrary function of the closed-string modulus. In fact, the generating function $G(t,\hat{t}_{1},\hat{t}_{2})$ is well defined up to an additive function of the closed-string modulus. However, $g(t)$ does not play any role in the connection matrices, as the derivatives of the generating function $G(t,\hat{t}_{1},\hat{t}_{2})$ with respect to the open moduli always appear in the connection matrices.
\par
Using (\ref{reltwo}), one can show that the generating function $G(t,\hat t_1,\hat t_2)$ satisfies the following identities
\begin{eqnarray}\label{relfour}
\overrightarrow{W}_{t \hat{t}_{1}}\cdot\eta\cdot\overrightarrow{W}_{t \hat{t}_{1}}-\overrightarrow{W}_{\hat{t}_{1} \hat{t}_{1}}\cdot\eta\cdot\overrightarrow{W}_{t t}-G_{\hat t_1\hat t_1}C&=&0\ ,\cr
\overrightarrow{W}_{t \hat{t}_{2}}\cdot\eta\cdot\overrightarrow{W}_{t \hat{t}_{2}}-\overrightarrow{W}_{\hat{t}_{2} \hat{t}_{2}}\cdot\eta\cdot\overrightarrow{W}_{t t}-G_{\hat t_2\hat t_2}C&=&0\ ,\cr
\overrightarrow{W}_{t \hat{t}_{1}}\cdot\eta\cdot\overrightarrow{W}_{t \hat{t}_{2}}-\overrightarrow{W}_{\hat{t}_{1} \hat{t}_{2}}\cdot\eta\cdot\overrightarrow{W}_{t t}-G_{\hat t_1\hat t_2}C&=&0\ .
\end{eqnarray}
In summary, the connection matrices are expressed in terms of the potentials $F$, $\overrightarrow{W}$ and $G$. All connection matrices are given in terms of the generating matrix $R$
\begin{eqnarray}\label{Mtii}
R(t,\hat{t}_{1},\hat{t}_{2})=
\left(
  \begin{array}{cccccccccc}
    0 & t & \hat t_1 & \hat t_2 & 0 & 0 & 0 & 0 & 0 & 0 \cr
  0 & 0 & 0 & 0 & F_{tt} & W^{(1)}_{t} & W^{(2)}_{t} & 0 & 0 & 0 \cr
  0 & 0 & 0 & 0 & 0 & W^{(1)}_{\hat{t}_{1}} & W^{(2)}_{\hat{t}_{1}} & 0 & 0 & 0 \cr
  0 & 0 & 0 & 0 & 0 & W^{(1)}_{\hat{t}_{2}} & W^{(2)}_{\hat{t}_{2}} & 0 & 0 & 0 \cr
  0 & 0 & 0 & 0 & 0 & 0 & 0 & t & G_{\hat t_1} & G_{\hat t_2} \cr
  0 & 0 & 0 & 0 & 0 & 0 & 0 & 0 & W^{(2)}_{\hat{t}_{1}}  & W^{(2)}_{\hat{t}_{2}}  \cr
  0 & 0 & 0 & 0 & 0 & 0 & 0 & 0 & W^{(1)}_{\hat{t}_{1}}  & W^{(1)}_{\hat{t}_{2}} \cr
  0 & 0 & 0 & 0 & 0 & 0 & 0 & 0 & 0 & 0 \cr
  0 & 0 & 0 & 0 & 0 & 0 & 0 & 0 & 0 & 0 \cr
  0 & 0 & 0 & 0 & 0 & 0 & 0 & 0 & 0 & 0
  \end{array}
\right)
  \ ,
\end{eqnarray}
where the second derivative of the prepotential $F_{tt}$ depends only on the closed-string modulus, while all the remaining entries are functions of both open- and closed-string moduli. Then the connection matrices arise as the gradient of $R$, namely
\begin{equation} \label{Mtiii}
\begin{aligned}
     M_t(t,\hat t_1,\hat t_2)&=\partial_t R(t,\hat t_1,\hat t_2) \ , \cr
     M_{\hat t_1}(t,\hat t_1,\hat t_2)&=\partial_{\hat t_1} R(t,\hat t_1,\hat t_2) \ , \cr
     M_{\hat t_2}(t,\hat t_1,\hat t_2)&=\partial_{\hat t_2} R(t,\hat t_1,\hat t_2) \ .
\end{aligned}
\end{equation}
This is the structure of flat connection matrices of the example with one closed- and two open-string moduli, and is compatible with the integrability requirement
\begin{equation}\label{curv}
  [\nabla_{t},\nabla_{\hat{t}_{1}}]=0\ ,\quad
  [\nabla_{t},\nabla_{\hat{t}_{2}}]=0\ ,\quad
  [\nabla_{\hat{t}_{1}},\nabla_{\hat{t}_{2}}]=0\ .
\end{equation}
This structure can be generalized for the more moduli cases as well.


\subsubsection{Quintic Example}
As we observed in section \ref{onetwomod}, the integrability requirement of the Gauss-Manin connection for the case of one closed- and two open-string deformations imposes several relations ((\ref{relone}), (\ref{reltwo}), (\ref{relthree}), and (\ref{relfour})) among the relative periods of the open-closed geometry. In this section, by taking an explicit example with one closed- and two open-string moduli and working out the complete set of linearly independent periods, we examine the imposed constraints on the periods of the open-closed system coming from the integrability requirement. In our example, the bulk geometry of the B-model is taken to be the mirror quintic. Similarly as in ref.\cite{Alim:2010za}, we introduce a two-parameter family of divisors defined in the mirror quintic by a polynomial of degree five. The two parameters in this family of divisors are identified with two open-string deformation parameters. More concretely, the defining equations of the bulk and boundary geometries are given by
\begin{eqnarray}
P&=&a_1\,x_{1}^{5}+a_2\, x_{2}^{5}+a_3\, x_{3}^{5}+a_4\, x_{4}^{5}+a_5\, x_{5}^{5}+a_0\, x_{1}x_{2}x_{3}x_{4}x_{5}\ ,\label{PandQone}\\
Q&=&b_1\, x_{1}^{5}+b_2\, x_{2}^{5}+b_0\,x_{1}x_{2}x_{3}x_{4}x_{5}\ ,\label{PandQtwo}
\end{eqnarray}
respectively. Our main goal is to verify the relations (\ref{relone}), (\ref{reltwo}), (\ref{relthree}), and (\ref{relfour}) by finding the corresponding set of relative periods. We notice that the family (\ref{PandQtwo}) contains the following holomorphic curves
\begin{align}
C^{\alpha}_{\pm}&=\{x_{2}+x_{3}=0\, ,x_{4}+x_{5}=0\, ,x_{1}^{2}\pm\sqrt{5\psi}\,x_{2}x_{4}=0\}\ ,\label{curvealpha}\\
C^{\beta}_{\pm}&=\{x_{1}+x_{3}=0\, ,x_{4}+x_{5}=0\, ,x_{2}^{2}\pm\sqrt{5\psi}\,x_{1}x_{4}=0\}\ ,\label{curvebeta}
\end{align}
by setting $b_{2}=0$ and $b_{1}=0$, respectively. Using these curves one can analyze the relation among on-shell and off-shell domain wall tensions as in ref.~\cite{Alim:2010za}.

We consider the charge vectors associated with the open-closed geometry
\begin{equation*}
\begin{array}{cccccccccc}
  & a_0 & a_1 & a_2 & a_3 & a_4 & a_5 & b_0 & b_1 & b_2 \\
 l_1: & -3 & 0 & 0 & 1 & 1 & 1 & -2 & 1 & 1 \\
 l_2: & -1 & 0 & 1 & 0 & 0 & 0 & 1 & 0 & -1 \\
 l_3: & 0 & 1 & -1 & 0 & 0 & 0 & 0 & -1 & 1
\end{array}
\end{equation*}
The good coordinates in the vicinity of a maximal unipotent point of monodromy are expressed in terms of the algebraic moduli in (\ref{PandQone}) and (\ref{PandQtwo}) via the following relations
\begin{eqnarray}\label{LVcoor}
z_{1}=\frac{a_3a_4a_5b_1b_2}{a_0^3b_0^2}\quad,\quad z_{2}=\frac{a_2b_0}{a_0b_2}\quad,\quad z_{3}=\frac{a_1b_2}{a_2b_1}\ ,
\end{eqnarray}
where the combination $z_{1}z_{2}^2z_{3}$ gives rise to the bulk large volume coordinate $z=\frac{a_1a_2a_3a_4a_5}{a_0^{5}}$.
\par
There are ten linearly independent solutions for the open-closed system. We have explicitly presented the solutions to the Picard-Fuchs system of differential operators in the vicinity of the large volume point after implementing the mirror map in appendix \ref{SolApp}.

Now, in order to examine the relations we found in the previous section, namely (\ref{relone}), (\ref{reltwo}), and (\ref{relthree}), we first need to identify the periods of the holomorphic three-form as appropriate linear combinations of the above solutions. For the closed-string sector, the prepotential and its derivative, $F_{0}$ and $F_{t}$, are identified by \cite{Candelas:1990rm,Klemm:1992tx}:
\begin{eqnarray}\label{Fzero}
F_{0}&=&\frac{5}{6}\,(\Pi_{7}+8\Pi_{8}+\Pi_{9})\ ,\cr
F_{t}&=&-\frac{5}{2}\,\Pi_{4}-10\,\Pi_{5}-\frac{5}{2}\,\Pi_{6}-\frac{21}{2}\,
(\Pi_{1}+2\Pi_{2}+\Pi_{3})\ .
\end{eqnarray}
The integral relative periods in the open-string sector can be determined most easily from the CY 4-fold associated to the brane geometry by open-closed duality \cite{Alim:2009rf,Alim:2009bx,Jockers:2009ti}. In particular $\eta$ arises as part of the intersection matrix of the associated Calabi-Yau fourfold geometry. For the sake of the integrability analysis the precise linear combination is not essential, however, and we simply make a choice for the superpotential periods. $W^{(1)}$ and $W^{(2)}$ are taken to be $W^{(1)}=\Pi_{4}$ and $W^{(2)}=\Pi_{5}$. For the $T^{(i)}$ periods, we choose the following linear combination:
\begin{eqnarray}\label{Tper}
\left(
  \begin{array}{c}
    T^{(1)} \cr T^{(2)}
  \end{array}
\right)=\left(
          \begin{array}{cc}
            \frac{2}{15} & \frac{2}{3}\cr \frac{1}{6} & \frac{4}{3}
          \end{array}
        \right)
\left(
  \begin{array}{c}
    \Pi_{7}\cr \Pi_{8}
  \end{array}
\right)
\ .
\end{eqnarray}
Then the metric $\eta$ introduced in previous subsection reads
\begin{eqnarray}\label{etamet}
\eta=\left(
       \begin{array}{cc}
         \frac{1}{10} & \frac{1}{4}\cr \frac{1}{4} & 1
       \end{array}
     \right)
\ .
\end{eqnarray}
With this identification, one finds that (\ref{relone}), (\ref{reltwo}),  (\ref{relthree}), and (\ref{relfour}) for the above example are indeed fulfilled. Therefore, the flat connection matrices of this example obey the general structure of (\ref{Mtii}),(\ref{Mtiii}) and (\ref{curv}).

\subsection{Application to non-Abelian branes}
As another interesting example we consider a special class of reducible divisors $\cx D$. More specifically we require $\cx D = \cx D_1 + \ldots +\cx D_N$, where the $N$ irreducible components $\cx D_\ell$ are defined by a single polynomial $p(x_i;\xi)=0$ which depends on $\ell$ only through the deformation parameter $\xi=\xi_\ell$. We will argue that the $N$ D7-branes associated to the components $\cxH_\ell$ enjoy an interpretation as $N$ parallel branes. For generic values of the open parameters $\xi_\ell$ the branes are separated and we describe the Coulomb branch of the worldvolume gauge theory. On the other hand if the deformation parameters of two or more components become the same, the corresponding brane components coincide and we expect gauge symmetry enhancement on the worldvolume of the D7-branes. Adding the appropriate flux, the relative periods compute the superpotential for non-Abelian D5-branes.

As a result the parameters $\xi_\ell$ may be interpreted as the eigenvalues of chiral multiplets $\Phi$ in the adjoint representation of $U(N)$.\footnote{We can decompose the multiplet $\Phi$ into the traceless parts and into singlet arising from the trace of the multiplet $\Phi$. Then the singlet captures the center of mass deformation of the parallel branes.} Then the superpotential induced from worldvolume fluxes, that is to say from lower-dimensional D5-brane charges, gives rise to the non-abelian superpotential
\begin{equation} \label{nonAbSup}
  W(\Phi) = \int {\rm tr}\ i_\Phi\Omega  \wedge F \ ,
\end{equation}
with the (non-abelian) worldvolume flux $F = dA + A \wedge A$. Here the multiplet $\Phi$ is an adjoint-valued section of the normal bundle with respect to the brane worldvolume, which is contracted with the holomorphic three form $\Omega$ of the Calabi-Yau threefold to the adjoint-valued two-form $i_\Phi\Omega$ \cite{Myers:1999ps}. The stated (non-abelian) superpotential is further discussed in refs.~\cite{Lust:2005bd,Jockers:2005zy,Gomis:2005wc,Martucci:2006ij,Alim:2009bx}, and it arises as the dimensional reduction of the non-abelian holomorphic Chern-Simons superpotential for branes filling the entire internal Calabi-Yau threefold \cite{Witten:1992fb,Kachru:2000an,Aganagic:2000gs}.

To illustrate these ideas we now discuss two particular examples, namely parallel branes on the mirror of the conifold and parallel branes on the mirror of the quintic threefold. In particular we exhibit the structure of the relative period vectors in flat coordinates.

\subsubsection{Parallel branes on the mirror conifold}
Our first example concerns parallel branes on the mirror of the conifold. This simple example allows us to study the relation of the chiral multiplet $\Phi$ to the position of the individual brane components $\xi_\ell$.

The mirror of the conifold is given in $\mathbbm{C}^4$ as the hypersurface \cite{Hori:2000kt}
$$
  x\,y=a_0\, \mathrm{e}^u+ a_1 \,\mathrm{e}^v+a_2\, \mathrm{e}^{u+v}+a_3\ ,
$$
which depends on the complex structure modulus $z=\frac{a_0a_1}{a_2a_3}$. The intersection of the above hypersurface equation with
\begin{equation} \label{defQcon}
  Q_N=b_0\,\mathrm{e}^{Nu} +b_1 \, \mathrm{e}^{(N-1)u+v}+ \ldots +b_N \,\mathrm{e}^{Nv}=0\ ,
\end{equation}
defines the family of divisors $\cxH$ modeling the parallel branes and depending on the open-string parameters $b_\ell$, which combine to $N$ open-string moduli fields. Note that the divisor $\cxH$ is reducible and splits into $N$ irreducible components $\cxH = \cx D_1 + \ldots +\cx D_N$. This can be seen by factorizing the defining divisor equation $Q_N$ into $N$ components
\begin{equation} \label{defQconfac}
  Q_N\,\sim\,\prod_{\ell=1}^N\,  ( \xi_\ell a_0 \mathrm{e}^u +a_1 \mathrm{e}^v  \,)\ ,
\end{equation}
where now the individual factors describe a single brane component $\cxH_\ell$ depending on the open-string field $\xi_\ell$. Note that the symmetry group $S_N$ acting on the fields $\xi_\ell$ and on the divisor components $\cxH_\ell$  leaving the defining equation \eqref{defQconfac} and hence the (reducible) divisor $\cxH$ invariant. Furthermore, viewing the individual components $\cxH_\ell$ as $N$ parallel branes, we interpret the symmetry group $S_N$ as the Weyl group of the $U(N)$ gauge theory, which acts on the position of the individual branes in the Coulomb phase of the $U(N)$ gauge theory. Therefore, from a gauge theory perspective the component fields $\xi_\ell$ represent the eigenvalues of the chiral matter multiplet $\Phi$, which in a suitable gauge reads
\begin{equation} \label{algmattercon}
  \Phi = {\rm Diag} \left( \xi_1 \, , \ldots \, , \xi_N \right) \ .
\end{equation}
Note that the parameters $b_\ell$ of the divisor equation \eqref{defQcon} are related to the eigenvalues $\xi_\ell$ of the matter multiplet $\Phi$ by
\begin{equation} \label{relbxi}
  \frac{b_k}{b_N}\left(\frac{a_1}{a_0}\right)^{N-k} = s_{N-k}(\xi_1,\ldots , \xi_N) \ ,
\end{equation}
in terms of the elementary symmetric polynomials in the variables $\xi_\ell$
\begin{equation} \label{sym}
  s_k(\xi_1,\ldots , \xi_N)=\sum_{1\le \ell_1< \ldots < \ell_k \le N} \xi_{\ell_1} \cdot \ldots \cdot \xi_{\ell_k} \ .
\end{equation}

The Picard-Fuchs differential equations of the hypergeometric system for the relative periods of the (reducible) divisor \eqref{defQcon} on the mirror conifold are determined in terms of the charge vectors
\begin{equation} \label{concharges}
\begin{array}{ccccccccccccccccc}
  &&& a_0 & a_1 & a_2 & a_3 & b_0 & b_1 & b_2 & b_3 & \cdots& b_{N-2}&b_{N-1}&b_{N}&& \\
  l&=&( & 1 & 1 & -1 & -1 & 0 & 0 & 0 & 0 &\cdots& 0 &0 & 0& )&\ ,\\
  \hat l_1&=&( & 0 & 0 & 0 & 0 & 1 & -2 & 1 & 0 & \cdots& 0 & 0 & 0& )&\ ,\\
  \hat l_2&=&( & 0 & 0 & 0 & 0 & 0 & 1 & -2 & 1 & \cdots& 0 & 0 & 0 & )&\ ,\\
  &&&&&\vdots&&&\vdots&&&\ddots&&\vdots\\
  \hat l_{N-1}&=&( & 0 & 0 & 0 & 0 & 0 & 0 & 0 & 0 & \cdots & 1 & -2 & 1& )&\ ,\\
  \hat l_{N}&=&( & -1 & 1 & 0 & 0 & 0 & 0 & 0 & 0 & \cdots& 0 & 1 & -1& )&\ .
\end{array}
\end{equation}
The charge vector $l$ captures the closed-string geometry of the bulk geometry, while the vectors $\hat l_1$ to $\hat l_N$ give rise to the matter fields associated to the $N$ parallel brane components. The corresponding algebraic coordinates $z, \hat z_1,\ldots , \hat z_N$ on the open-closed moduli space read
$$
  z = \frac{a_0 a_1}{a_2a_3} \ , \qquad
  \hat z_\ell = \frac{b_{\ell-1} b_{\ell+1}}{b_\ell^2} \ , \  \ell = 1,\ldots, N-1 \ , \qquad
  \hat z_N = \frac{a_1 b_{N-1}}{a_0 b_N} \ .
$$

It is straightforward to see that all the algebraic coordinates $\hat z_\ell$ can be expressed in terms of symmetric polynomials $s_k$ with respect to the eigenvalues $\xi_\ell$. As a consequence the algebraic coordinates $\hat z_\ell$ are gauge invariant quantities expressible in terms of the adjoint-valued matter multiplet $\Phi$.

Solving the Picard-Fuchs system of differential equations reveals in the vicinity of $(z,\hat z_\ell)=0$ of the open-closed moduli space a regular (constant) solution together with $N+1$ logarithmic solutions, namely
$$
  t(z) = \frac{1}{2\pi i} \log z \ , \quad \hat t_\ell(\hat z_k) = \frac{1}{2\pi i}\log \left(\hat z_\ell \cdots \hat z_N\right) + p_\ell(\hat z_k) \ , \ \ell=1,\ldots,N \ ,
$$
where $p_\ell(\hat z_k)$ are holomorphic functions in the vicinity $(z,\hat z_\ell)=0$. Here, $t(z)$ and $\hat t_\ell(z_k)$ are the closed- and open-string flat coordinates, respectively.

Since the flat coordinates $\hat t_\ell$ are functions of the gauge invariant algebraic coordinates $\hat z_\ell$, they can again be expressed in a manifest $U(N)$-gauge theoretic manner in terms of the matter multiplet $\Phi$. A careful examination of the Picard-Fuchs system \eqref{concharges} reveals that the coordinates $\hat t_\ell$ assemble themselves into a flat adjoint-valued matter multiplet $\hat{\bold t}$, which is given in terms of the matrix relation
\begin{equation} \label{flatmattercon}
 \hat{\bold t}(\Phi) \,=\, \frac{1}{2\pi i}\log \Phi \ .
\end{equation}
In particular in the gauge \eqref{algmattercon} we find
$$
  \hat{\bold t} \,=\,  {\rm Diag} \left( \hat t_1 \, , \ldots \, , \hat t_N \right) \ .
$$

To illustrate the discussed structures we now consider two simple explicit examples. Clearly, for a single brane the adjoint-valued matter field $\hat{\bold t}$ becomes the (gauge-invariant) single flat coordinate $\hat t_1 = \tfrac{1}{2\pi i}\log \hat z_1 = \tfrac{1}{2\pi i}\log \xi_1$ capturing the deformation modulus of the irreducible divisor $\cxH \equiv \cxH_1$. For two parallel branes, {\it i.e.} for $N=2$, the Picard-Fuchs differential operators yield two open-string flat coordinates
\begin{equation}
  \hat t_{1/2}\,=\,\frac{1}{2\pi i}\log\left(\frac{\hat z_2}{2}(1\pm\sqrt{1-4\hat z_1})\right)
  \,=\, \frac{1}{2\pi i}\log \left(\frac{1}{2}\, {\rm tr}\,\Phi \pm\frac{1}{2} \sqrt{\left({\rm tr}\,\Phi\right)^2 - 4 \det \Phi} \right) \ .
\end{equation}
The arguments of the logarithm on the right hand side are the eigenvalues of the algebraic matter multiplet $\Phi$, and hence we identify the flat coordinates as \begin{equation}
\hat t_{1/2} = \frac{1}{2\pi i} \log \xi_{1/2}\, .\end{equation} The two flat coordinates $\hat t_1$ and $\hat t_2$ describe now the flat deformation moduli associated to the two components $\cxH_1$ and $\cxH_2$ of the reducible divisor $\cxH = \cxH_1 + \cxH_2$. Note that the structure of the flat $U(2)$-matter multiplet $\hat{\bold t}={\rm Diag}\,(\hat t_1, \hat t_2)$ is in agreement with the general expression \eqref{flatmattercon} for the flat $U(N)$-matter multiplet.

As side remark let us briefly point out the relation of the discussed open-closed conifold geometry to its dual A-model fourfold description as studied in refs.~\cite{Mayr:2001xk,Alim:2009rf,Alim:2009bx}. The dual A-model fourfold is a fibration of the non-compact conifold threefold over a disk with a singular central fiber. This degeneration is semi-stable and the fourfold itself is smooth. The bulk flat coordinate $t$ measures the (quantum) volume of the generic conifold fiber, whereas the open flat coordinates $\hat t_\ell$ determine the (quantum) volume of holomorphic curves in the central fiber. As $1<k\le N$ of the flat open coordinates $\hat t_\ell$ coincide, {i.e.} $\hat t_{i_1}=\hat t_{i_2}= \ldots = \hat t_{i_k}$ (with mutually distinct indices $i_n$)  a $A_{k-1}$-singularity arises at the central fiber. The appearance of the singularity signals a symmetry enhancement, which arises in the dual brane picture from coinciding brane components $\cxH_{i_1}\equiv\ldots\equiv\cxH_{i_k}$. This gauge symmetry enhancement of theories with four supercharges encoded in the singularity structure of the fourfolds geometry is similar to the gauge symmetry enhancement of theories with eight supercharges arising from Calabi-Yau threefold singularities \cite{Katz:1996ht,Klemm:1996kv}.

\subsubsection{Parallel branes on the mirror quintic}
As our next example we discuss the parallel branes on the mirror quintic. The mirror quintic arises as the Calabi-Yau hypersurface
$$
  P\,=\,a_1\, x_1^5+a_2\, x_2^5+a_3\, x_3^5+a_4\, x_4^5+a_5\, x_5^5+a_{0}\, x_{1}x_{2}x_{3}x_{4}x_{5}= 0 \ .
$$
in the projective space $\IP^4$ orbifolded by the Greene-Plesser group $\IZ_5^3$, which acts by appropriate phase rotations on the homogeneous coordinates $x_\ell$ of the projective space $\IP^4$. The complex structure modulus $z$ is given in terms of the coefficients $a_0$ to $a_5$ by $z=\frac{a_1\cdots a_5}{a_0^5}$.

We realize the parallel brane components in terms of the (reducible) divisor $\cxH = \cxH_1 + \ldots + \cxH_N$ by the degree $4N$ homogeneous equation
\begin{equation} \label{Qparquintic}
  Q_N \,=\,  \sum_{k=0}^N b_k \, x_1^{4k}\,(x_2x_3x_4x_5)^{N-k} \ .
\end{equation}
The parameters $b_0,\ldots , b_N$ encode the open-string deformations of the parallel branes. By factorizing the polynomial \eqref{Qparquintic} the irreducible components of the divisor $\cxH$ become manifest, namely
\begin{equation} \label{xiquin}
  Q_N \,\sim\, \prod_{\ell=1}^N (\, \xi_\ell \, a_0 \, x_2x_3x_4x_5 + a_1 \, x_1^4 \,)\ , \qquad
   \frac{b_k}{b_N}\left(\frac{a_1}{a_0}\right)^{N-k} = s_{N-k}(\xi_1,\ldots , \xi_N) \ ,
\end{equation}
in terms of the symmetric polynomials \eqref{sym}. As in the previous example the factorized form exhibits the symmetry with respect to the symmetric group $S_N$ acting on the open-string deformation parameters $\xi_\ell$. Interpreting the $S_N$ symmetry as the Weyl group of the underlying $U(N)$ gauge theory, we construct the chiral matter multiplet $\Phi$, which takes in terms of the open string deformation parameters the form
\begin{equation} \label{algMatQuin}
  \Phi = {\rm Diag}\,(\xi_1\,, \ldots , \xi_N) \ .
\end{equation}

The brane geometry in the large complex structure phase of the mirror quintic is governed by the charge vectors
\begin{equation} \label{chargeparquin}
\begin{array}{ccccccccccccccccccc}
  &&& a_0 & a_1 & a_2 & a_3 & a_4 & a_5 & b_0 & b_1 & b_2 & b_3 & \cdots& b_{N-2}&b_{N-1}&b_{N}&& \\
  l_0&=&( & -4 & 0 &   1 & 1 & 1 & 1 & -1 & 1 & 0 & 0 &\cdots& 0 &0 & 0& )&\ ,\\
  l_1&=&( & 0 & 0 & 0 & 0 & 0 & 0 & 1 & -2 & 1 & 0 & \cdots& 0 & 0 & 0& )&\ ,\\
  l_2&=&( & 0 & 0 & 0 & 0 & 0 & 0 & 0 & 1 & -2 & 1 & \cdots& 0 & 0 & 0 & )&\ ,\\
  &&&&\vdots&&&\vdots&&&\vdots&&&\ddots&&\vdots&&&\\
  l_{N-1}&=&( & 0 & 0 & 0 & 0 & 0 & 0 & 0 & 0 & 0 & 0 & \cdots & 1 & -2 & 1& )&\ ,\\
  l_N&=&( & -1 & 1 & 0 & 0 & 0 & 0 & 0 & 0 & 0 & 0 & \cdots& 0 & 1 & -1& )&\ ,
\end{array}
\end{equation}
and the vertices of the polyhedron
\begin{equation} \label{qinpoly}
\begin{aligned}
  (0,0,0,0,0)\,,\;(1,0,0,0,0)\,,\;(0,1,0,0,0)\,,\;&(0,0,1,0,0)\,,\;(0,0,0,1,0)\,,\;(-1,-1,-1,-1,0)\,,\\ &(n,0,0,0,1)\,;\;\;\;\;\;\; 0\leq n\leq N\,. \end{aligned}
\end{equation}
Note that adding a brane component simply corresponds to adding an additional vertex to the polyhedron. The algebraic open-closed coordinates arising from the charges \eqref{chargeparquin} read
\begin{equation} \label{zquin}
  z_0\,=\, \frac{a_2a_3a_4a_5b_1}{a_0^4b_0} \ , \quad
  z_k\,=\, \frac{b_{k-1}b_{k+1}}{b_k^2} \ , \ k=1,\ldots,N-1 \ , \quad
  z_{N} \,=\, \frac{a_1 b_{N-1}}{a_0b_N} \ ,
\end{equation}
and the algebraic bulk complex structure modulus is given by
\begin{equation} \label{bulkcoord}
  z = z_0 \cdots z_N \ .
\end{equation}
In the following we also work with the coordinates $(z,\hat z_\ell)$, which are comprised of the bulk coordinate $z=z_0$ and the open-string coordinates $\hat z_\ell \equiv z_\ell$, $\ell=1,\ldots, N$, as the coordinates $\hat z_\ell$ relate directly to the open-string deformation parameters $\xi_\ell$.

The large complex structure phase of a single (irreducible) brane component is discussed in detail ref.~\cite{Alim:2009bx}. In summary the fundamental period together with the logarithmic periods are given by
$$
\begin{aligned}
 \Pi_0^{N=1}(z_0,z_1) \,&=\, \Pi_{gen}^{N=1}(z_0,z_1,0,0) \ , \\
 \Pi_{1}^{N=1}(z_0,z_1) \,&=\, \frac{1}{2\pi i} \left.\partial_{\rho_0}\Pi_{gen}^{N=1}(z_0,z_1,\rho_0,0)\right|_{\rho_0=0} \ , \\
 \Pi_{2}^{N=1}(z_0,z_1) \,&=\, \frac{1}{2\pi i} \left.\partial_{\rho_1}\Pi_{gen}^{N=1}(z_0,z_1,0,\rho_1)\right|_{\rho_1=0} \ ,
\end{aligned}
$$
in terms of the generating functional
$$
  \Pi_{gen}^{N=1}(z_0,z_1,\rho_0,\rho_1)\,=\, \sum_{n_0,n_1=0}^{+\infty}
   \frac{\Gamma(4(n_0+\rho_0)+(n_1+\rho_1)+1)}{\Gamma(n_0+\rho_0+1)^4\Gamma(n_1+\rho_1+1)}z_0^{n_0+\rho_0}
   z_1^{n_1+\rho_1} \ ,
$$
which (due to eq.~\eqref{bulkcoord}) yields the flat closed coordinate $t$ and flat open coordinates $\hat t$
\begin{equation}\label{teehatone}
\begin{aligned}
  t(z_0,z_1) \,&=\, \frac{\Pi_1^{N=1}(z_0,z_1)+\Pi_2^{N=1}(z_0,z_1)}{\Pi_0^{N=1}(z_0,z_1)}
  \,=\, \frac{1}{2\pi i} \log z_0z_1 + \ldots \ , \\
  \hat t(z_0,z_1) \,&=\, \frac{\Pi_2^{N=1}(z_0,z_1)}{\Pi_0^{N=1}(z_0,z_1)}
  \,=\, \frac{1}{2\pi i} \log z_1 + \ldots \ .
\end{aligned}
\end{equation}

Then in terms of these flat coordinates the whole relative period vector reads \cite{Alim:2009bx}
$$
  \vec\Pi_{flat}^{N=1}(t,\hat t) \,=\, \left( 1 , t, \hat t, F_t(t), W(t,\hat t), -F_0(t), T(t,\hat t) \right) \ ,
$$
in terms of the bulk prepotential $F(t)$, the superpotential $W(t,\hat t)$, and the top period in the open string sector $T(t,\hat t)$.

Analogously to the branes on the mirror conifold, the generalization to $N$ brane components, as described by the reducible divisor \eqref{Qparquintic}, replaces the single flat open-coordinate $\hat t$ by a flat adjoint-valued matter multiplet $\hat{\bold t}$. As in eq.~\eqref{algMatQuin}, it has the diagonal structure in terms of the flat coordinates of the individual brane components
$$
 \hat{\bold t}(z,\Phi) \,=\,
  \begin{pmatrix}
    \tfrac{\log (\hat z_1\cdots \hat z_N)}{2\pi i} +\ldots &&& \\
   &\ddots&&\\
   && \tfrac{\log (\hat z_{N-1}\hat z_N)}{2\pi i}+\ldots \\
   &&&\tfrac{\log \hat z_N}{2\pi i}+\ldots
  \end{pmatrix} \ .
$$
Then the superpotential $W$ and the top period $T$ become $U(N)$ adjoined-valued periods $\bold W$ and $\bold T$ as functions of the flat adjoined-valued coordinates $\hat{\bold t}$, and we obtain the relative (adjoined-valued) period vector as a function of the adjoint-valued matter multiplet $\hat{\bold t}$
\begin{equation} \label{flatquinvec}
  \vec{\bold\Pi}_{flat}^{N}(t,\hat{\bold t}) \,=\, \left( 1 , t,\hat{\bold t}, F_t(t), \bold W(t,\hat{\bold t}), -F_0(t),\bold T(t,\hat{\bold t}) \right) \ .
\end{equation}
However, since we only calculate the diagonal components of the adjoined-valued entries in this relative period vector, we cannot unambiguously extend $W$ to $\bold W$ and $T$ to $\bold T$ by this procedure. For instance, similarly as in ref.~\cite{Myers:1999ps}, we expect the appearance of commutators involving $\hat{\bold t}$, which are not visible as long as $\hat {\bold t}$ is diagonal. It would be interesting to explicitly compute such intrinsically non-abelian contributions to make contact with the non-abelian nature of the superpotential~\eqref{nonAbSup}.

We now illustrate the described generalization to $N$ brane components by constructing the flat relative period vector for two parallel brane components. That is to say we analyze the open-closed deformation problem for the reducible divisor $\cxH = \cxH_1 + \cxH_2$ given by the divisor equation $Q_{N=2}$. Then the ten linearly independent solutions to the system of Picard-Fuchs equations yield ten hypergeometric relative periods, which are generated by the functional
$$
\begin{aligned}
 &\Pi_{gen}^{N=2}(z_0,z_1,z_2,\rho_0,\rho_1,\rho_2)\,=\, \\
 &\quad\sum_{n_k=0}^{+\infty}\!\!
   \tfrac{\Gamma(4(n_0+\rho_0)+(n_1+\rho_1)+1)\,z_0^{n_0+\rho_0}z_1^{n_1+\rho_1} z_2^{n_2+\rho_2}}
   {\Gamma(n_0+\rho_0+1)^4\Gamma(n_2+\rho_2+1)\Gamma(n_1+\rho_1-(n_0-\rho_0)+1)\Gamma(n_1+\rho_1-(n_2-\rho_2)+1)\Gamma(n_0+\rho_0-2(n_1-\rho_1)+n_2+\rho_2+1)}
    \ .
\end{aligned}
$$
This yields the relevant periods for constructing the flat coordinates
$$
\begin{aligned}
 \Pi_0^{N=2}(z_0,z_1,z_2) \,&=\, \Pi_{gen}^{N=2}(z_0,z_1,z_2,0,0,0) \ , \\
 \Pi_{k}^{N=2}(z_0,z_1,z_2) \,&=\, \frac{1}{2\pi i} \left.\partial_{\rho_{k-1}}\Pi_{gen}^{N=1}(z_0,z_1,z_2,\rho_0,\rho_1,\rho_2)\right|_{\rho=0} \ , \ k=1,2,3 \ .
\end{aligned}
$$
The flat coordinates become
\begin{equation}\label{flattwo}
\begin{aligned}
  t(z_0,z_1,z_2) \,&=\, \tfrac{\Pi_1^{N=2}(z_0,z_1,z_2)+\Pi_2^{N=2}(z_0,z_1,z_2)+\Pi_3^{N=2}(z_0,z_1,z_2)}{\Pi_0^{N=2}(z_0,z_1,z_2)}
  \,=\, \frac{1}{2\pi i} \log z_0z_1z_2 + \ldots \ , \\
  \hat t_1(z_0,z_1,z_2) \,&=\, \frac{\Pi_3^{N=2}(z_0,z_1,z_2)+\Pi_3^{N=2}(z_0,z_1,z_2)}{\Pi_0^{N=2}(z_0,z_1,z_2)}
  \,=\, \frac{1}{2\pi i} \log z_1z_2 + \ldots \ , \\
  \hat t_2(z_0,z_1,z_2) \,&=\, \frac{\Pi_3^{N=2}(z_0,z_1,z_2)}{\Pi_0^{N=2}(z_0,z_1,z_2)}
  \,=\, \frac{1}{2\pi i} \log z_2 + \ldots \ .
\end{aligned}
\end{equation}
Alternatively, we can also compute the flat coordinates $\hat t_1$ and $\hat t_2$ by using the diagonal gauge of the algebraic matter multiplet $\Phi\,=\,{\rm Diag}(\xi_1, \xi_2)$. According to eqs.~\eqref{xiquin} and \eqref{zquin} the open-string component deformation parameters are given in terms of the algebraic coordinates $z_1$ and $z_2$ as
$$
  \xi_{1/2} \,=\,  \frac{1}{2} z_2 \left(1 \pm \sqrt{1-4 z_1} \right) \ .
$$
Inserting $\xi_1$ and $\xi_2$ into the single brane open-string flat coordinate $\hat t$, we indeed find the expected relationship
$$
  \hat t_{1/2}(z,\hat z_1,\hat z_2) \equiv \hat t(z, \xi_{1/2}(\hat z_1,\hat z_2)) \ ,
$$
where  $\hat t_{1/2}$ represent the flat coordinates of the two parallel brane components and $\hat t$ the flat coordinate single brane component in eq.~\eqref{teehatone}. Furthermore, in the diagonal gauge and in flat coordinates the remaining six solutions to the Picard-Fuchs differential equations split into two double-logarithmic superpotential solutions and two triple-logarithmic top period solutions, obeying $W^{N=2}_{1/2}(t,\hat t_1,\hat t_2)\equiv W^{N=1}(t,\hat t_{1/2})$ and $T^{N=2}_{1/2}(t,\hat t_1,\hat t_2)\equiv T^{N=1}(t,\hat t_{1/2})$ , and the two flat bulk periods $F_t(t)$ and $-F_0(t)$. Thus for two brane components we have explicitly confirmed (in the diagonal gauge) the general structure of the period vector \eqref{flatquinvec}.

Analogously to the branes on the mirror conifold, as defined by the charge vectors \eqref{chargeparquin} and the polyhedron \eqref{qinpoly}, the open-closed relative periods for parallel branes on the mirror quintic enjoy again a dual A-model formulation on the dual Calabi-Yau fourfold \cite{Mayr:2001xk,Alim:2009rf}.

More generally, for a mirror Calabi-Yau threefold $Z^*$, given as a hypersurface in a toric ambient space, together with a reducible divisor $\cxH$ representing $N$ parallel brane components defined by the homogeneous equation
\begin{equation}
  Q_N\,=\,b_0 X_a^N+ b_1 X_a^{N-1}X_b+ ...+b_N X_b^N \, \sim \, \prod_{\ell=1}^N \xi_\ell X_a + X_b \ ,
\end{equation}
we obtain a dual A-model fourfold formulation. Here $X_a$ and $X_b$ represent some monomials, which appear in the defining hypersurface equation of the Calabi-Yau threefold $Z^*$, and the parameters $b_0$ to $b_N$, or alternatively the brane component deformation parameters $\xi_\ell$, encode the open-string deformation of the $N$ brane components. Then the Calabi-Yau fourfold geometry $X$, which arises as described in refs.~\cite{Mayr:2001xk,Alim:2009rf,Alim:2009bx} and as exemplified for the mirror quintic in eqs.~\eqref{chargeparquin} and \eqref{qinpoly}, is realized as a (non-compact) hypersurface in a toric ambient space. This non-compact Calabi-Yau fourfold is the Calabi-Yau threefold $Z$, which is the mirror to the threefold $Z^*$, fibered over a disk. The central threefold fiber $Z$ over disk degenerates semi-stably such that the Calabi-Yau fourfold $X$ is smooth, and the structure of the central fiber encodes the geometry of the brane components.

As part of the toric construction of the fourfold hypersurface $X$ we add $N+1$ vertices of the form $(n (\nu_a-\nu_b),1)$, where $n$ runs from $0$ to $N$ and the vertices $\nu_a$ and $\nu_b$ are associated to the monomials $X_a$ and $X_b$ in the toric description of the threefold $Z^*$. We notice that these vertices span the Dynkin diagram of $A_{N-1}$ and the dual Calabi-Yau fourfold develops an $A_{k}$-singularity, $1<k\le N-1$, as $k$ of the deformation parameters $\xi_\ell$ agree. Hence the interplay of the gauge symmetry enhancement arising from coinciding parallel branes translates on the dual Calabi-Yau fourfold into the emergence of singularities. This is expected in view of the dual M-theory picture developed in refs.~\cite{Aganagic:2009jq,Mertens:2011ha}. It would be interesting to make the gauge symmetry enhancement of parallel brane components and the relation to singularities in the dual fourfold formulation more precise, and we plan to come back to this issue elsewhere.


\section{Summary and conclusions}

In this paper, we have studied the flatness and integrability structure of the B-model Gauss-Manin system for open-closed geometries with several deformations. We have shown how these conditions allow to define distinguished flat coordinates and the superpotential function at an arbitrary point in the open-closed deformation space. As an application, we have studied Gromow-Witten invariants at different limit points of the open-closed deformation space for a brane in local $\mathbbm{P}^2$.

It was previously shown \cite{Alim:2009bx} that for the simplest example with one closed- and one open-string deformation, the flatness of the Gauss-Manin connection is related to the K3-structure of the subsystem of the open-closed geometry. However, the open-closed subsystem does not generically exhibit a K3-structure, when several open- and closed-string deformations are considered. From the integrability of the Gauss-Manin system, we have extracted necessary and sufficient conditions among the relative periods of the open-closed geometry in full generality. These are non-linear relations among the derivatives of the relative periods written in flat coordinates. It is worth mentioning that the derived relations are globally valid throughout the open-closed moduli space and not specific to a certain regime. In the limit where only one closed- and one open-string deformation survive, it is shown that these relations descend to the previous case, where flatness of the Gauss-Manin connection is guaranteed by the K3 structure of the subsystem.
\par
Furthermore, in order to examine the general integrability relations that we have discovered, we have provided explicit examples. In these examples, we have computed the full set of relative periods in a corner of the open-closed deformation space and after expressing the periods in flat coordinates, we have found that the general integrability relations are indeed fulfilled.

Along the way we have uncovered certain interesting properties of the explicit examples we have studied. In particular, we have explained in our last example how to realize a system of parallel branes in the compact setup and how to capture the superpotential associated with its deformations through the program of the variation of mixed Hodge structure.
We have studied how the non-abelian gauge symmetry on the world-volume of a stack of parallel branes develops, as one varies the moduli that are present in the problem. It would be very interesting to extend these computations to correlation functions with boundary changing operators, and we hope to come back to these questions in the future.


\acknowledgments
We would like to thank Vincent Bouchard, Andrea Brini, Ilka Brunner and Albrecht Klemm for discussions and comments.
M.A. is supported by the DFG fellowship AL 1407/1-1.
The work of M.H. and P.M. is supported by the program ``Origin and Structure of the Universe" of the German Excellence Initiative and the Deutsche Forschungsgemeinschaft.
The work of H.J. was partially supported by the Stanford Institute of Theoretical Physics and the NSF Grant 0244728 and partially by the Kavli Institute for Theoretical Physics and the NSF Grant PHY05-51164.
The work of A.M. is supported by the Studienstiftung des deutschen Volkes. The work of M.S. was supported by a EURYI award of the European Science Foundation during the completion of this project at Arnold Sommerfeld Center for Theoretical Physics in Munich.
M.A. thanks the KITP for hospitality during completion of this work, this was supported in part by DARPA under Grant No.
HR0011-09-1-0015 and by the NSF Grant PHY05-51164.
H.J. and M.S. are also supported by the DFG grant KL 2271/1-1.

\appendix

\section{Appendix }
{\it Solutions of P.F. operators for the two open-string moduli example}\label{SolApp}\\[1mm]
There are ten linearly independent solutions for the example of the mirror quintic with the two-parameter family of divisors and they are organized in the following way. There are three single $\log$ solutions which basically define the mirror maps
\begin{eqnarray}\label{logone}
\Pi_{0}=1\quad,\quad \Pi_{1}=\log(q_{1})\quad,\quad \Pi_{2}=\log(q_{2})\quad,\quad \Pi_{3}=\log(q_{3})\ .
\end{eqnarray}
There exist three double $\log$ solutions which give rise to the two superpotential periods and the closed-string derivative of the prepotential of the closed sector. They are given by
\begin{eqnarray}\label{logtwo}
\Pi_{4}&=&\log ^2\left(q_1\right)+1035 q_1^2+\frac{5 q_2^2}{2}+60 q_1-240 q_1 q_2+10 q_2+10 q_2 q_3\cr
&&+\frac{10}{9} \left(31704 q_1^3-5589
   q_2 q_1^2+162 q_2^2 q_1-216 q_2 q_3 q_1+q_2^3\right)\ ,\\
   \Pi_{5}&=&\log ^2\left(q_2\right)+\log \left(q_1\right) \log \left(q_2\right)-15 q_1-4 q_2+q_3+\frac{1}{4} \left(-1035 q_1^2+204 q_2 q_1-4 q_2^2+q_3^2-4 q_2 q_3\right)\cr
   && +\frac{1}{9} \left(-79260 q_1^3+12879 q_2 q_1^2-1233 q_2^2 q_1+621 q_2 q_3 q_1-4
   q_2^3+q_3^3\right)\ ,\\
   \Pi_{6}&=& \log ^2\left(q_3\right)+2 \log \left(q_1\right) \log \left(q_3\right)+4 \log \left(q_2\right) \log
   \left(q_3\right)+\frac{2 q_2^3}{3}+368 q_1 q_2^2+486 q_1^2 q_2\cr
   && -q_3^2+\frac{3 q_2^2}{2}+36 q_1 q_2-36 q_1 q_3 q_2-6 q_3 q_2+6 q_2-\frac{4
   q_3^3}{9}-4 q_3\ ,
\end{eqnarray}
Finally there exist three triple $\log$ solutions which correspond to the prepotential of the closed sector and the two top periods $T^{(1)}$ and $T^{(2)}$:
\begin{eqnarray}\label{logthree}
\Pi_{7}&=&\log ^3\left(q_1\right)+105680 q_1^3 \log \left(q_1\right)+\frac{10}{3} q_2^3 \log \left(q_1\right)+3105 q_1^2 \log
   \left(q_1\right)+540 q_1 q_2^2 \log \left(q_1\right)\cr
   && +\frac{15}{2} q_2^2 \log \left(q_1\right)+180 q_1 \log
   \left(q_1\right)-18630 q_1^2 q_2 \log \left(q_1\right)-720 q_1 q_2 \log \left(q_1\right)+30 q_2 \log \left(q_1\right)\cr
   && -720 q_1 q_2 q_3 \log \left(q_1\right)+30 q_2 q_3 \log \left(q_1\right)+\frac{80620 q_1^3}{3}-\frac{160 q_2^3}{9}-\frac{405
   q_1^2}{2}+225 q_1 q_2^2-\frac{45 q_2^2}{2}\cr
   && -180 q_1-5130 q_1^2 q_2+540 q_1 q_2+60 q_2-15 q_2^2 q_3+540 q_1 q_2 q_3+60 q_2 q_3\ ,\\ \cr\cr
   \Pi_{8}&=&\log ^3\left(q_2\right)+\frac{3}{2} \log \left(q_1\right) \log ^2\left(q_2\right)-\frac{1}{2} q_2^3 \log
   \left(q_2\right)+\frac{1}{3} q_3^3 \log \left(q_2\right)+\frac{3}{4} \log ^2\left(q_1\right) \log \left(q_2\right)\cr
   && -276 q_1 q_2^2 \log \left(q_2\right)-\frac{9}{8} q_2^2 \log \left(q_2\right)+\frac{3}{4} q_3^2 \log \left(q_2\right)-\frac{729}{2}
   q_1^2 q_2 \log \left(q_2\right)-27 q_1 q_2 \log \left(q_2\right)\cr
   && -\frac{9}{2} q_2 \log \left(q_2\right)+27 q_1 q_2 q_3 \log
   \left(q_2\right)+\frac{9}{2} q_2 q_3 \log \left(q_2\right)+3 q_3 \log \left(q_2\right)-13210 \log \left(q_1\right)
   q_1^3\cr
   && -\frac{20155 q_1^3}{6}-\frac{2}{3} \log \left(q_1\right) q_2^3+\frac{35 q_2^3}{9}+\frac{1}{6} \log \left(q_1\right)
   q_3^3-\frac{4 q_3^3}{9}-\frac{3105}{8} \log \left(q_1\right) q_1^2+\frac{405 q_1^2}{16}\cr
   && -\frac{3}{2} \log \left(q_1\right)q_2^2-\frac{411}{2} \log \left(q_1\right) q_1 q_2^2+\frac{105}{2} q_1 q_2^2+\frac{45 q_2^2}{8}+\frac{3}{8} \log
   \left(q_1\right) q_3^2-\frac{9}{4} q_2 q_3^2-\frac{9 q_3^2}{16}\cr
   && -\frac{45}{2} \log \left(q_1\right) q_1+\frac{45 q_1}{2}+\frac{4293}{2} \log \left(q_1\right) q_1^2 q_2+\frac{4023}{4} q_1^2 q_2-6 \log \left(q_1\right) q_2+\frac{153}{2}
   \log \left(q_1\right) q_1 q_2\cr
   && -\frac{81 q_1 q_2}{2}-3 q_2+3 q_2^2 q_3+\frac{3}{2} \log \left(q_1\right) q_3-\frac{3}{2} \log
   \left(q_1\right) q_2 q_3+\frac{207}{2} \log \left(q_1\right) q_1 q_2 q_3\cr
   && -\frac{135}{2} q_1 q_2 q_3-\frac{15 q_2 q_3}{2}+\frac{3 q_3}{2}\ ,\\ \cr\cr
   \Pi_{9}&=& \log ^3\left(q_3\right)+3 \log \left(q_1\right) \log ^2\left(q_3\right)+6 \log \left(q_2\right) \log ^2\left(q_3\right)+3 \log
   ^2\left(q_1\right) \log \left(q_3\right)\cr
   && +12 \log ^2\left(q_2\right) \log \left(q_3\right)+12 \log \left(q_1\right) \log
   \left(q_2\right) \log \left(q_3\right)+2 \log \left(q_1\right) q_2^3+4 \log \left(q_2\right) q_2^3\cr
   && -\frac{4}{3} \log \left(q_1\right) q_3^3-\frac{8}{3} \log \left(q_2\right) q_3^3+\frac{32 q_3^3}{9}+\frac{9}{2}
   \log \left(q_1\right) q_2^2+9 \log \left(q_2\right) q_2^2+1104 \log \left(q_1\right) q_1 q_2^2\cr
   && +2208 \log \left(q_2\right) q_1 q_2^2-645 q_1 q_2^2-\frac{45 q_2^2}{2}-3 \log \left(q_1\right) q_3^2-6 \log \left(q_2\right) q_3^2-\frac{40 q_2^3}{3}+18 q_2 q_3^2\cr
   && +\frac{9 q_3^2}{2}+1458 \log \left(q_1\right) q_1^2 q_2+2916 \log \left(q_2\right) q_1^2 q_2-2916 q_1^2 q_2+18 \log \left(q_1\right) q_2\cr
   && +36 \log \left(q_2\right) q_2+108 \log \left(q_1\right) q_1 q_2+216 \log \left(q_2\right) q_1 q_2-216 q_1 q_2-36 q_2-9 q_2^2 q_3\cr
   && -12 \log \left(q_1\right) q_3-24 \log \left(q_2\right) q_3-18 \log \left(q_1\right) q_2 q_3-36 \log
   \left(q_2\right) q_2 q_3-108 \log \left(q_1\right) q_1 q_2 q_3\cr
   && -216 \log \left(q_2\right) q_1 q_2 q_3-12 q_3\ .
\end{eqnarray}

\end{document}